\documentclass[12pt,onecolumn,draftcls]{IEEEtran}
\pdfoutput=1
\usepackage[pdftex]{graphicx}

\usepackage{epsfig}

\usepackage{subfigure}

\usepackage{amssymb}

\usepackage{amsbsy}

\usepackage{amsmath}

\usepackage{cite}

\usepackage{float}

\usepackage{bbm}

\usepackage{url}

\usepackage{multirow}

\newtheorem{theorem}{Theorem}
\newtheorem{lemma}{Lemma}
\newtheorem{remark}{Remark}

\newtheorem{proposition}{Proposition}
\newtheorem{corollary}{Corollary}
\begin{document}
%
\title{On the Multiple Access Channel with Asynchronous Cognition}

\author{Michal Yemini$^{*}$, Anelia Somekh-Baruch$^{*}$ and Amir Leshem$^{*}$

\thanks{This research was partially supported by Israel Science Foundation (ISF) grant  2013/919.}
\thanks{The results of this paper were partially presented in \cite{us_ISIT_2014}.}
\thanks{$*$ Faculty of Engineering, Bar-Ilan University, Ramat-Gan, 52900, Israel. Email:
michal.yemini.biu@gmail.com, anelia.somekhbaruch@gmail.com, leshem.amir2@gmail.com.}
}
\date{}

\maketitle
\begin{abstract}
In this paper we introduce the two-user asynchronous cognitive multiple access channel (ACMAC). This channel model includes two transmitters, an uninformed one, and an informed one which knows prior to the beginning of a transmission the message which the uninformed transmitter is about to send. We assume that the channel from the uninformed transmitter to the receiver suffers a fixed but unknown delay. We further introduce a modified model, referred to as the asynchronous \textit{codeword} cognitive multiple access channel (ACC-MAC), which differs from the ACMAC in that the informed user knows the signal that is to be transmitted by the other user, rather than the message that it is about to transmit. We state  inner and outer bounds on the ACMAC and the ACC-MAC capacity regions, and we specialize the results to the Gaussian case.  Further, we characterize the capacity regions of these channels in terms of multi-letter expressions.
 Finally, we provide an example which instantiates the difference between message side-information and codeword side-information.
\end{abstract}

\IEEEpeerreviewmaketitle

\section{Introduction}
In recent years, due to the scarcity of free static spectrum resources, a new concept coined as ``Cognitive Radio'' \cite{Mitola,Haykin,GoldsmithJafar2009} has emerged. "Cognitive radio networks" may refer to several models and setups, however, generally speaking, the common assumption for the different interpretations of this term is the existence of users that can sense their surroundings and are able to change their configurations accordingly, these users are referred to as cognitive users. Though the knowledge that the cognitive users may  acquire about the network may vary from one model network to the other, the common goal is to improve spectrum utilization by giving the opportunity to more users to transmit while limiting their interference on non-cognitive users in the network. Further, in some models, the cognitive users can even help the non-cognitive users to improve their reliable communication rates. One possible model of cognition assumes that the cognitive users possess knowledge of the codewords or messages that licensed users transmit. The information theoretic analysis of these cognitive models is closely related to the Gel'fand Pinsker channel \cite{GP} and the cognitive MAC \cite{GoldsmithJafar2009,Devroye}, hence our motivation to further broaden our knowledge of these channel models.

Channels with side-information at the transmitter have been widely studied from the information-theoretic perspective. One of the earliest models was introduced by Shannon \cite{Shannon}. In \cite{Shannon}, Shannon analyzed the point-to-point  state-dependent memoryless channel with causal side-information at the transmitter, and established a single-letter formula for its capacity. Another well known model is the point-to-point state-dependent memoryless channel with noncausal side-information at the transmitter, which is also known as the Gel'and-Pinsker (GP) channel. The capacity of this channel was found in \cite{GP}.
Side-information also plays a role in multi-user channels, such as the multiple access channel (MAC). The capacity region of the discrete memoryless MAC was found in terms of a multi-letter expression in \cite{vanderMeulen1971} and was further characterized by a single-letter expression in \cite{Ahlswede1971}.  The MAC with correlated sources is analyzed in \cite{SlepianWolf1973}.  In this channel model each transmitter has two messages that it wishes to send, a private message and a common message which both transmitters share. The capacity region of this channel is achieved by superposition techniques as described in \cite{SlepianWolf1973}. For other related models see \cite{Anelia2008,LiSimeone2013,ZaidiPiantanida2013,BrossLapidoth2012,MaricYates2007}.

The classical MAC model assumes that the channel is synchronous, however, this is not necessarily the case in practical channels. Several extensions of the MAC to the asynchronous setup have been studied, see e.g.,  \cite{CoverMcEliece1981,HuiHumblet1985,Verdu1989}. It was shown
that the capacity region of the discrete memoryless MAC depends on the nature of the delay that may occur in the channel. It was shown
\cite{CoverMcEliece1981} that if the delay is finite or grows slowly relatively to the block length, the capacity region remains the same as if there is no delay in the channel. Hui and Humblet \cite{HuiHumblet1985} proved that the capacity region may be smaller if the delay is of the same order of the block length, since time sharing cannot be used. Asynchronism in MAC with memory was considered by  Verd\'{u} \cite{Verdu1989b} under the assumption that the asynchronism is not bounded.

\begin{figure}[H]
  \centering
  \includegraphics[scale=0.45]{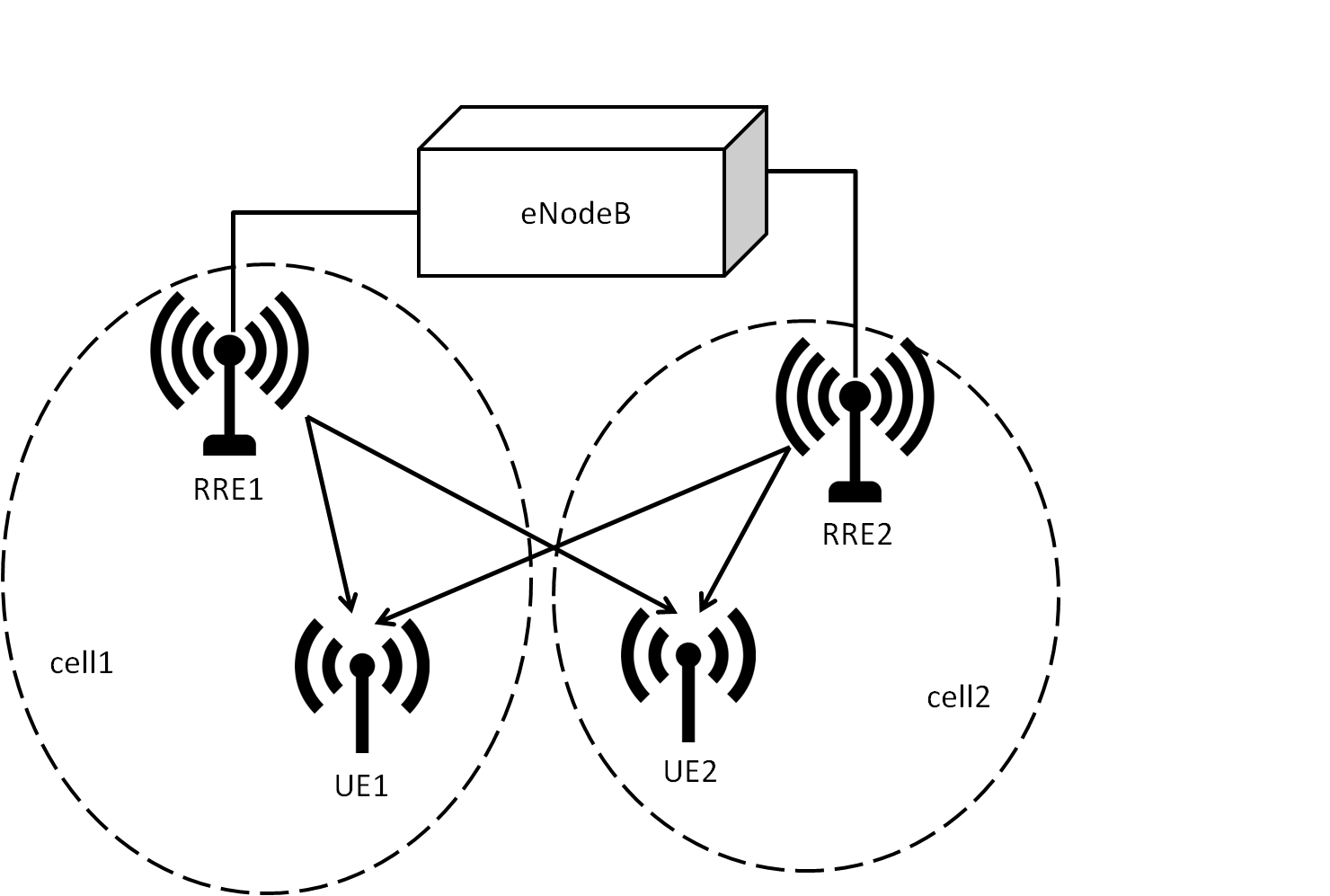}\\
  \caption{An asynchronous CoMP \cite{LiuLi2013}}\label{comp_fig}
\end{figure}

Channels with side-information at the transmitter may also assume that this side-information is synchronized with the channel. However, this assumption is not always realistic and asynchronism is present in many practical communication systems. One example for such a practical setups is a cellular network in which coordinated multipoint (CoMP) techniques are used (see for example \cite{Karakayali2006,Irmer2011}). These techniques can be used in the downlink and can involve different schemes for cooperation and coordination of base-stations. Additionally, one can also take advantage of base-stations cooperation in the uplink, for example several base-stations can jointly decode received signals. While in optimal scenarios all the cooperative nodes are synchronized there can be synchronization issues in these schemes (see for example \cite{Irmer2011}). An example for asynchronous CoMP is discussed in \cite{LiuLi2013}, in this setup two remote radio equipments (RREs) serve two user equipments (UEs) via joint transmissions. The two RREs are connected to the same eNodeB by high quality optical fiber channels (therefore, no delays are present in these channels). It is assumed that the channels from the RREs to the UEs suffer from random time offsets due to continuously varying multi-path environment. This scheme is depicted in Fig.~\ref{comp_fig} which is presented in \cite{LiuLi2013}.

In previous works, \cite{Eilat_us,us_paper1}, we analyzed several point-to-point state dependent channels with asynchronous side information.
In this paper, we inspect how asynchronous knowledge affects the performance of cognitive multi-user channels.
The asynchronous cognitive MAC  is composed of a receiver and two transmitters, an uninformed transmitter that wishes to send a message, and a cognitive one which is informed of the other transmitter's message and/or codeword. It is assumed that the MAC is asynchronous, that is, the channel from the uninformed user to the receiver suffers an unknown but bounded delay. We characterize the capacity region of this channel in terms of a multi-letter expression, and state  inner and outer bounds on its capacity region. In this paper, we consider two variations of the asynchronous cognitive MAC.
The first setup we consider is the asynchronous cognitive MAC with {\it message-cognition} at one transmitter (ACMAC) \cite{us_ISIT_2014}. An additional setup we consider is the asynchronous \textit{codeword} cognitive MAC  (ACC-MAC), depicted in Fig.~\ref{ACC-MAC_fig}. The difference between the ACC-MAC model and the ACMAC model is that in the former, the informed encoder knows prior to transmission the uninformed encoder's codeword, whereas in the latter model it knows the uninformed encoder's message and consequently its codeword. Thus, by definition, in the ACMAC model the informed encoder can send some of the information bits of the uninformed user's message, whereas in the ACC-MAC model this is no longer possible. Consequently, the capacity region of the ACC-MAC is contained in the capacity region of the ACMAC, where the inclusion is usually strict.  We note that the results of this paper were partially presented in \cite{us_ISIT_2014}.

The rest of the paper is organized as follows. Section~\ref{sec:Channel Models} includes several notations and definitions and also the ACMAC and ACC-MAC models. Section~\ref{sec:Achievable ACMAC} states the capacity region of the ACMAC in terms of a multi-letter expression and also includes  inner and outer bounds on its capacity region. In Section~\ref{sec:Examples_ACMAC} we address the Gaussian ACMAC and state inner and outer bounds on its capacity region.  Further, Section~\ref{ACC-MAC_sec} presents the capacity region of the ACC-MAC in terms of a multi-letter expression and additionally establishes inner and outer bounds on its capacity region. In Section~\ref{sec:example} we present an example for a channel in which the ACC-MAC's capacity region is strictly smaller than the ACMAC's  capacity region. Finally, Section~\ref{sec:Conclusion} concludes the paper.

\section{Notations, Definitions, and Assumptions}\label{sec:Channel Models}
We use the following notations and definitions:
A vector $(x_1,\ldots,x_n)$ is denoted by the boldface notation $\textbf{x}$, whereas the vector  $(x_i,\ldots,x_j)$ is denoted by $x_i^j$. In the special case in which $\textbf{x}$ is a vector whose $i$-th entry, $x_i$, is not a scalar but a vector,  the notation $x_{i,j}$ signifies the $j$ entry of $x_i$. In certain cases, the vector $\textbf{x}$ is denoted by $x^n$ as well.
The probability law of a random variable $X$ is denoted by $P_X$ and $\mathcal{P}(\mathcal{X})$ denotes the set of distributions on the alphabet $\mathcal{X}$.
Further, $\mathbbm{1}_{\{A\}}$ denotes the indicator function, i.e., $\mathbbm{1}_{\{A\}}$ equals $1$ if the statement $A$ holds and $0$ otherwise.
We also denote the closure of a subset, $A$, of a metric space by $\text{closure}(A)$.

For simplicity of the presentation, throughout this paper, we assume that the set of possible delays in the asynchronous cognitive multiple access channels is
$\mathcal{D}=\{-d_{min},-d_{min}+1,\ldots,d_{max}\}$, where $0 \leq d_{min},d_{max}$, it follows that $D=d_{max}+d_{min}+1$. Additionally, throughout this paper we assume that all transmitters and receivers know a-priori the (finite) values $d_{min}$ and $d_{max}$. Further, we assume that the delay $d$ is fixed during the transmission of a codeword over the channel. We note that the results which are derived in this paper can be easily generalized to arbitrary finite sets of delays, i.e., sets of  finite numbers that are not necessarily sequential numbers. Moreover, the results in this paper also hold in the general case, in which the delay is randomly distributed over a finite set and then is fixed during the transmission of a codeword.

The set of all $n$ vectors $x^n\in\mathcal{X}^n$ that are $\epsilon$-strongly typical \cite[p.\ 326]{CT} with respect to $P_X\in {\cal P}({\cal X})$ is denoted by $T_{\epsilon}^n(X)$. Additionally, we denote  by $T_{\epsilon}^n(X|y^n)$ the set of all $n$ vectors $x^n$ that are $\epsilon$-strongly jointly typical with the vector $y^n$ with respect to a probability mass function (p.m.f.) $P_{X,Y}$.
Further, let $P_{X,Y}$ be a conditional p.m.f. from $\mathcal{X}$ to $\mathcal{Y}$. For $x\in\mathcal{X}$ denote by $\{P_d(y|x)\}_{d\in\mathcal{D}}$ a set of conditional p.m.f.'s from ${\cal X}$ to $\cal Y$,  that depend on the value of $d$. Let $d\in\mathcal{D}$ and let $x,y$ be two random variables with a set of joint p.m.f.'s $\{P_d(x,y)\}_{d\in\mathcal{D}}$, for each value of $d$ we denote the information theoretic functionals of the respective p.m.f. $P_d(x,y)$ by the subscript $d$, e.g., $I_d(X;Y)$.
Finally, for each value $d\in\cal D$  we use the notation $T_{d,\epsilon}^n(X,Y)$ to denote the set of all $\epsilon$-strongly jointly typical sequences in $\mathcal{X}^n\times \mathcal{Y}^n$ with respect to the p.m.f. $P_d(x,y)$.

\subsection {The Asynchronous Cognitive Multiple Access Channel Model}\label{ACMAC_def}

The cognitive multiple access channel (MAC) is a stationary discrete memoryless multiple access channel which is defined by the channel input alphabets $\mathcal{X}_1$ and $\mathcal{X}_2$, the channel output alphabet $\mathcal{Y}$, and the channel transition probabilities $P(y|x_1,x_2)$. The CMAC model assumes a unidirectional knowledge where the transmitter of user $2$ knows in advance the message of user $1$ (the uninformed user), and consequently its codeword too.

\begin{figure}[H]
\centering
\includegraphics{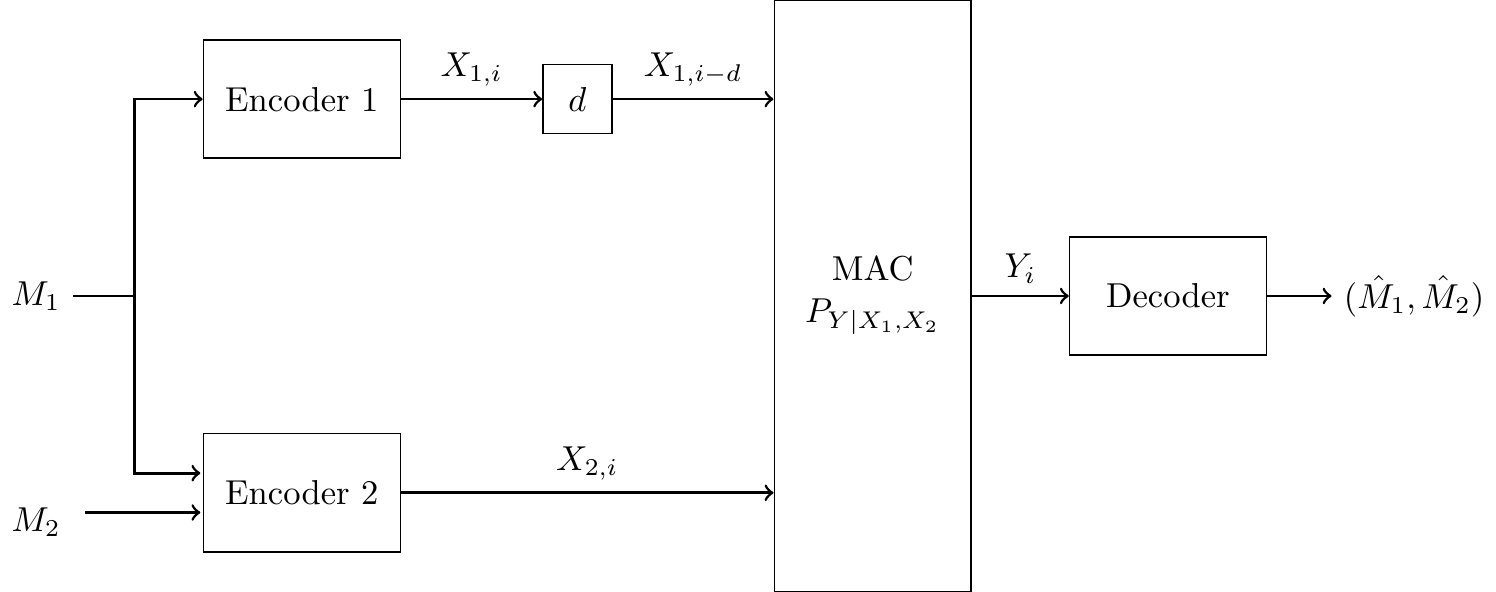}\\
\caption{Asynchronous cognitive multiple access channel.}
\label{ACMAC_fig}
\end{figure}

The asynchronous cognitive multiple access channel (ACMAC), which is depicted in Fig.\ \ref{ACMAC_fig}, is a CMAC  with  a delay $d\in\mathcal{D}$ between the encoder of the uninformed user and the channel. It is assumed that this delay is fixed during the transmission of a codeword.

Under these assumptions, the channel transition probabilities from  $\mathcal{X}_1^n \times \mathcal{X}_2^n$ to $\mathcal{Y}^n$ are defined by
\begin{flalign}\label{trans_prob_ACMAC}
P_d(\textbf{y} |\textbf{x}_1,\textbf{x}_2)=\prod_{i=1}^n P_{Y|X_1,X_2}(y_i|x_{1,i-d},x_{2,i}),\quad d\in\mathcal{D}.
\end{flalign}
Additionally, for all $i\in \{1,\ldots,n\}$ such that $i-d\notin\{1,\ldots,n\}$,  $x_{1,i-d}$ are arbitrary.
{
   \def\OldComma{,}
    \catcode`\,=13
    \def,{%
      \ifmmode%
        \OldComma\discretionary{}{}{}%
      \else%
        \OldComma%
      \fi%
    }%
Let $\mathcal{M}_1=\{1,2,\ldots,2^{nR_1}\}$, and $\mathcal{M}_2=\{1,2,\ldots,2^{nR_2}\}$, and assume that the messages $M_1$ and $M_2$ are independent random variables uniformly distributed over the sets $\mathcal{M}_1$ and $\mathcal{M}_2$, respectively.
A $(2^{nR_1},2^{nR_2},n)$-code for the ACMAC channel consists of the deterministic encoding functions
\begin{flalign}
&f_{1,n}:\mathcal{M}_1\rightarrow\mathcal{X}_1^n \\
&f_{2,n}:\mathcal{M}_1\times \mathcal{M}_2 \rightarrow\mathcal{X}_2^n
 \end{flalign}
and a deterministic decoding function
\begin{flalign}
&g_n:\mathcal{Y}^n \rightarrow \mathcal{M}_1 \times \mathcal{M}_2.
\end{flalign}
Define the average probability of error for $d\in\mathcal{D}$ as
\begin{flalign}
\bar{P}_{e,d} &=  \frac{1}{2^{nR_1}2^{nR_2}} \sum_{m_1=1}^{2^{nR_1}}\sum_{m_2=1}^{2^{nR_2}} \sum_{\textbf{y}: g_n(\textbf{y})\neq (m_1,m_2)} P_d\left(\textbf{y} |f_{1,n}(m_1),f_{2,n}(m_1,m_2) \right),
\end{flalign}
where $P_d\left(\textbf{y} |f_{1,n}(m_1),f_{2,n}(m_1,m_2) \right)$ is defined by (\ref{trans_prob_ACMAC}), and for all $i\in \{1,\ldots,n\}$ such that $i-d\notin\{1,\ldots,n\}$,  $x_{1,i-d}$ are arbitrary.

A $(2^{nR_1},2^{nR_2},n)$-code is said to be a $(2^{nR_1},2^{nR_2},n,\epsilon)$-code, if $\bar{P}_{e,d}\leq\epsilon$ for all $d \in\mathcal{D}$. A rate-pair $(R_1,R_2)$ is said to be achievable for the ACMAC channel, if there exists a sequence of $\left(2^{nR_1},2^{nR_2},n,\epsilon_n \right) $-codes with $\epsilon_n \rightarrow 0$ as $n \rightarrow \infty$.
\newline
The capacity region of the ACMAC is defined as the closure of the set of all achievable rate-pairs.}

\subsection {The Channel Model of the Asynchronous Cognitive Multiple Access Channel with Codeword Knowledge at One Encoder}\label{sec:ACC-MAC_def}

\begin{figure}[H]
\centering
\includegraphics{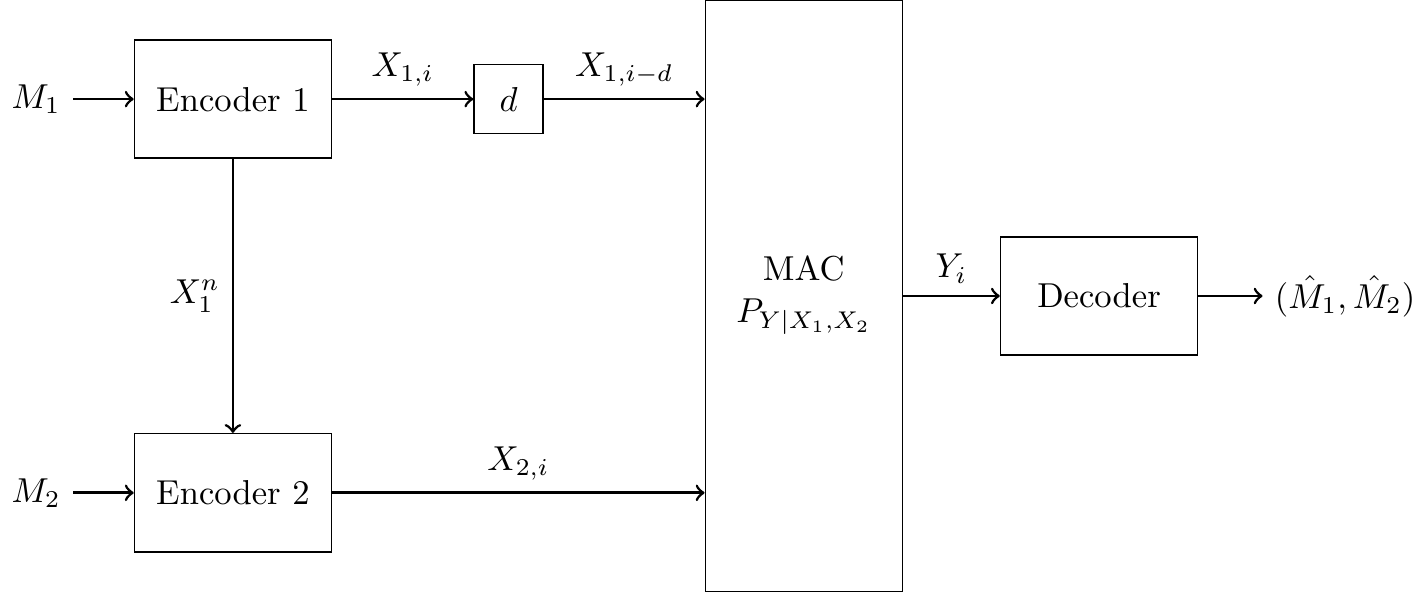}\\
\caption{Asynchronous cognitive MAC with codeword knowledge at one encoder.}
\label{ACC-MAC_fig}
\end{figure}

The definitions for the asynchronous \textit{codeword} cognitive MAC  (ACC-MAC) are similar to those of the ACMAC, with the following modification: We consider unidirectional knowledge where transmitter $2$ knows the {\it codeword} of user $1$ (the uninformed user) prior to the beginning of transmission. This difference can be seen in Fig.\ \ref{ACC-MAC_fig} which depicts the  ACC-MAC.

{
   \def\OldComma{,}
    \catcode`\,=13
    \def,{%
      \ifmmode%
        \OldComma\discretionary{}{}{}%
      \else%
        \OldComma%
      \fi%
    }%

The encoder of the informed user is therefore defined by the deterministic mapping:
\begin{flalign}
&f_{2,n}: \mathcal{X}_1^n \times\mathcal{M}_2\rightarrow\mathcal{X}_2^n, \label{Eq:encoder2}
 \end{flalign}
and the average probability of error takes on the form:
\begin{flalign}
\bar{P}_{e,d} &=  \frac{1}{2^{n(R_1+R_2)}} \sum_{m_1=1}^{2^{nR_1}}\sum_{m_2=1}^{2^{nR_2}} \sum_{\textbf{y}: g_n(\textbf{y})\neq (m_1,m_2)} P_d\left(\textbf{y} |f_{1,n}(m_1),f_{2,n}(f_{1,n}(m_1),m_2) \right),
\end{flalign}
where $P_d\left(\textbf{y} |f_{1,n}(m_1),f_{2,n}(f_{1,n}(m_1),m_2) \right)$ is defined in (\ref{trans_prob_ACMAC}).
The capacity region of the ACC-MAC is defined similarly to that of the ACMAC.}

We emphasize that in the ACMAC model the informed encoder knows both the message and the codeword of the uninformed encoder, that is, $m_1$ and $f_{1,n}(m_1)$. In the ACC-MAC model, as opposed to the ACMAC model, the informed encoder knows the uninformed encoder's codeword $f_{1,n}(m_1)$, but not necessarily its message $m_1$.
It follows that in the ACMAC model the mapping between the message to the codeword need not be reversible, that is, two or more messages can share the same codeword in the codebook of user 1. In the ACMAC scenario, the uninformed encoder relies on the informed encoder to transmit the remaining information. However, similar coding scheme cannot be adopted in the ACC-MAC model since the informed encoder knows the uninformed encoder's codeword $f_{1,n}(m_1)$, but not necessarily its message $m_1$.

\subsection{Known Delay at the Receiver}
In  the above channel models, i.e., ACMAC and ACC-MAC, we assume that the decoder does not know the actual delay in the channel before decoding the message. However, since the set of delays $\mathcal{D}$ is finite, by sending a predetermined training sequence in the first $o(n)$ bits, the decoder can deduce the delay with probability of error that vanishes as $n$ tends to infinity. Thus, we omit the transmission of the training sequence in our coding schemes and assume hereafter that the decoder knows the delay $d$ prior to the decoding stage.

\section{Bounds on the Capacity Region of the ACMAC }\label{sec:Achievable ACMAC}

 This section is devoted to the ACMAC (see Fig.~\ref{ACMAC_fig} and Section~\ref{ACMAC_def}). We present the capacity region of the ACMAC in terms of a multi-letter expression
  and derive  outer and inner bounds on its capacity region.
\subsection{A Multi-letter Expression for the Capacity Region of the ACMAC}\label{ACMAC_capacity_sec}
Even though multi-letter expressions are usually not tractable, they can yield significant results and in certain cases even computable formulae  (see for example \cite{MotahariKhandani2009,Shang2009,AnnapureddyVeeravalli2009}).  We next provide a multi-letter formula for the capacity region of the ACMAC.

Denote by $\textbf{X}_1$, $\textbf{X}_2$ the codewords $(X_{1,1},X_{1,2},\ldots,X_{1,n})$ and $(X_{2,1},X_{2,2},\ldots,X_{2,n})$ of users 1 and 2, respectively.
Additionally, let $d\in\mathcal{D}$, and let $P_d(\textbf{x}_1,\textbf{x}_2,\textbf{y})=P(\textbf{x}_1,\textbf{x}_2)P_d(\textbf{y}|\textbf{x}_1,\textbf{x}_2)$ where $P_d(\textbf{y}|\textbf{x}_1,\textbf{x}_2)$ is defined in (\ref{trans_prob_ACMAC}).
Define the region of rate-pairs $(R_1,R_2)$
\begin{flalign}\label{R_n_def_ACMAC}
\mathcal{R}_n=\bigcup_{P(\textbf{x}_1,\textbf{x}_2)}\bigcap_{d\in\mathcal{D}}
\left\{
\begin{array}{ll}
    \multirow{2}{*}{$(R_1,R_2):$}& R_1 +R_2 \leq \frac{1}{n}I_d(\textbf{X}_1,\textbf{X}_2;\textbf{Y}),\\
    &\hspace{1cm}R_2 \leq \frac{1}{n}I_d(\textbf{X}_2;\textbf{Y}|\textbf{X}_1)
\end{array}
\right\},
\end{flalign}
and additionally define the region
\begin{flalign}
\mathcal{Q}_n=\bigcup_{P(\textbf{x}_1,\textbf{x}_2)}\bigcap_{d\in\mathcal{D}}\left\{
\begin{array}{ll}
    \multirow{2}{*}{$(R_1,R_2):$}& R_1 +R_2 \leq \frac{1}{n}I_{d}(\textbf{X}_1,\textbf{X}_2;Y_{d_{max}+1}^{n-d_{min}}),\\
    &\hspace{1cm}R_2 \leq \frac{1}{n}I_{d}(\textbf{X}_2;Y_{d_{max}+1}^{n-d_{min}}|\textbf{X}_1)
\end{array}
\right\}
\end{flalign}
where
$P_{d}(y_{d_{max}+1}^{n-d_{min}}|\textbf{x}_1,\textbf{x}_2)=\prod_{i=d_{max}+1}^{n-d_{min}}P(y_i|x_{1,i-d},x_{2,i})$.

Recall that, as noted in Section~\ref{sec:Channel Models}, $D=|{\cal D}|$, we next derive the capacity region of the ACMAC in terms of a multi-letter expression.
\begin{theorem}\label{ACMAC_capacity}
Let $P_{Y|X_1,X_2}$ be the channel transition probability of an ACMAC with a finite set of possible delays $\mathcal{D}$. The capacity region of the ACMAC is given by
\begin{flalign}\label{ACMAC_capacity_term}
\mathcal{C}=\text{closure}\left(\bigcup_{n \geq D} \mathcal{Q}_n\right)=
\text{closure}(\lim\sup_{n\rightarrow\infty}\mathcal{R}_n)
=\text{closure}(\lim\inf_{n\rightarrow\infty}\mathcal{R}_n).
\end{flalign}
\end{theorem}
The proof of this theorem appears in Appendix~\ref{ACMAC_C1}.

\begin{corollary}
Let $d\in\mathcal{D}$. The capacity region of the ACMAC is not affected by the transition probability of the first $d_{max}$ and last $d_{min}$ symbols.
\end{corollary}

Let $\mathcal{C}_{ACMAC}$ and $\mathcal{C}_{CMAC}$ be the capacity regions of the ACMAC and its corresponding CMAC (in which $\mathcal{D}=\{0\}$), respectively. In addition, let $\mathcal{C}_{MAC}$ be the capacity region of the asynchronous MAC with no cognition \cite{Alswede1971}. Since bounded/finite asynchronization  does not affect the capacity region of the MAC \cite{CoverMcEliece1981}, another conclusion which follows from Theorem \ref{ACMAC_capacity} is that as expected, $\mathcal{C}_{MAC}\subseteq\mathcal{C}_{ACMAC}\subseteq \mathcal{C}_{CMAC}$.

We note that the capacity region of the ACMAC is closed and convex. The region is closed by definition, and the convexity follows from standard arguments of time sharing between two codebooks  operating at two different rate-pairs.

\subsection{Inner and Outer Bounds on the Capacity Region of the ACMAC}\label{ACMAC_bounds}
This section presents inner and outer bounds on the capacity region of the ACMAC. To state these bounds we first  introduce     the random vector $\overline{V}$ and the probability function $P_d(x_1,\overline{v},x_2,y)$  which are relevant for the next sections as well. Let
\begin{flalign}\label{ACMAC_Vn}
\overline{V}= (V_1,\ldots,V_D)
\end{flalign}
where $V_{i}\in \mathcal{X}_1$.  The sequence  $\overline{V}$  acts as the vector of all input possibilities of the uninformed encoder to the channel at a time instant. Additionally, for $P_{X_1}\in\mathcal{P}(\mathcal{X}_1)$ define the following probability measures
\begin{flalign}
&P_{\overline{V}}(\overline{v})=\prod_{j=1}^{D}P_{X_1}(v_{j}) \label{PV_X1_ACMAC}.
\end{flalign}
Further, let
\begin{flalign}\label{acmac_single_def}
P_d(x_1,\overline{v},x_2,y)=P(\overline
{v})\mathbbm{1}_{\{x_1=v_{d_{max}-d+1}\}}P(x_2|\overline{v})P(y|x_1,x_2),
\end{flalign}
where $\mathbbm{1}_{\{\cdot\}}$ is the indicator function which is defined in Section \ref{sec:Channel Models}.
We next present an achievable region for the ACMAC.

\begin{theorem}\label{ACMAC_inner_single}
Let $P_{Y|X_1,X_2}$ be an ACMAC with a finite set of possible delays $\mathcal{D}$.
Let $(X_1,\overline{V},X_2,Y)$ be distributed according to  (\ref{acmac_single_def}).
Denote,
\begin{flalign}\label{inner_region_ACMAC}
\underline{\mathcal{R}}&=\bigcup_{P(x_1),P(x_2|\overline{v})}
\left\{
\begin{array}{ll}
    \multirow{2}{*}{$(R_1,R_2):$}& R_1 +R_2 \leq \min_{d\in\mathcal{D}}\left[I_d(X_1;Y)+I_d(X_2;Y|\overline{V})\right],\\
    &\hspace{1cm}R_2 \leq \min_{d\in\mathcal{D}}I_d(X_2;Y|\overline{V})
\end{array}
\right\}.
\end{flalign}
The  closure convex  of $\underline{\mathcal{R}}$ is an achievable rate region for the ACMAC.
\end{theorem}

The proof of this theorem appears in Appendix~\ref{ACMAC_A2}.

Theorem \ref{ACMAC_inner_single} states that convex closure of the region (\ref{inner_region_ACMAC}) is an achievable rate region for the ACMAC. However, we acknowledge the fact that using time-sharing in  the expressions of the inequalities of (\ref{inner_region_ACMAC})  yields a larger
achievable rate-region. We next articulate this point.
Let $Q$ be some random variable with some probability function $P_Q$. Define for $P_{X_1}\in\mathcal{P}(\mathcal{X}_1)$ the following probability functions,
\begin{flalign}
&P_{\overline{V}|Q}(\overline{v}|q)=\prod_{j=1}^{D}P_{X_1}(v_{j}|q), \label{PV_X1_ACMAC}
\end{flalign}
and
\begin{flalign}\label{acmac_single_def_q}
&P_d(q,x_1,\overline{v},x_2,y)=P(q)P(\overline
{v}|q)\mathbbm{1}_{\{x_1=v_{d_{max}-d+1}\}}P(x_2|\overline{v},q)P(y|x_1,x_2).
\end{flalign}
\begin{corollary}
Let $P_{Y|X_1,X_2}$ be an ACMAC with a finite set of possible delays $\mathcal{D}$.
Let $(Q,X_1,\overline{V},X_2,Y)$ be distributed according to  (\ref{acmac_single_def_q}).
Denote,
\begin{flalign}\label{inner_region_ACMAC_q}
\underline{\mathcal{R}}&=\bigcup_{P(q),P(x_1|q),P(x_2|\overline{v},q)}
\left\{
\begin{array}{ll}
    \multirow{2}{*}{$(R_1,R_2):$}& R_1 +R_2 \leq \min_{d\in\mathcal{D}}\left[I_d(X_1;Y|Q)+I_d(X_2;Y|\overline{V},Q)\right],\\
    &\hspace{1cm}R_2 \leq \min_{d\in\mathcal{D}}I_d(X_2;Y|\overline{V},Q)
\end{array}
\right\}.
\end{flalign}
The  closure   of $\underline{\mathcal{R}}$ is an achievable rate region for the ACMAC.
\end{corollary}

An outer bound on the capacity region of the ACMAC is presented next. Let, \begin{flalign}\label{eq:dep_over_X1}
\overline{X}_1=(X_{1,1},\ldots,X_{1,D})
\end{flalign}
 and
\begin{flalign}\label{eq:dep_over_X2}
\overline{X}_2=(X_{2,1},\ldots,X_{2,D}).
\end{flalign}
Also, let
\begin{flalign}\label{eq:def_tilde_v}
\tilde{V}=(\overline{V}_1,\ldots,\overline{V}_D),
\end{flalign}
 where
\begin{flalign}\label{eq:def_bar_vi}
\overline{V}_i=(\overline{V}_{i,1},\ldots,\overline{V}_{i,D}),
\end{flalign}
and $\overline{V}_{i,j}\in\mathcal{X}_1$ for all $i,j\in \{1,\ldots,D\}$.
Further, let $Q$ be a random variable with a probability function $P_Q$ on some finite alphabet and define
\begin{flalign}\label{eq:v_bar_vec_outer}
P(\tilde{v}|q)=P(\overline{v}_1|q)\prod_{i=2}^D(\overline{v}_{i,D}|q,\overline{v}_{i-1},\ldots\overline{v}_1)\mathbbm{1}_{\{(\overline{v}_{i,1},\ldots,\overline{v}_{i,D-1})=(\overline{v}_{i-1,2},\ldots,\overline{v}_{i-1,D})\}},
\end{flalign}
and
\begin{flalign}\label{eq:outer_prob}
&P_d(q,\overline{x}_1,\tilde{v},\overline{x}_2,\overline{y}) =P(q)P(\tilde{v}|q)P(\overline{x}_1|\tilde{v})P(\overline{x}_2|\tilde{v},q)P_{d}(\overline{y}|\overline{x}_2,\overline{x}_1),\nonumber\\
&P_d(\overline{x}_1|\tilde{v})=\prod_{i=1}^{D}\mathbbm{1}_{\{x_{1,i}=\overline{v}_{i,d_{max}-d+1}\}},\nonumber\\
&P_d(\overline{x}_2|\tilde{v},q)=\prod_{i=1}^{D}P(x_{2,i}|x_{2}^{i-1},\tilde{v}),\nonumber\\
&P_{d}(\overline{y}|\overline{x}_2,\overline{x}_1)= \prod_{i=1}^D P_{d}(y_i|x_{2,i},x_{1,i}).
\end{flalign}

\begin{theorem}\label{ACMAC_outer_single}
Let $P_{Y|X_1,X_2}$ be an ACMAC with a finite set of possible delays $\mathcal{D}$. Denote,
\begin{flalign}\label{outer_region_ACMAC}
\overline{\mathcal{R}}&=\bigcup_{P(q),P(\tilde{v}|q),P(\overline{x}_2|\tilde{v},q)}
\left\{
\begin{array}{ll}
    \multirow{2}{*}{$(R_1,R_2):$}&R_1+R_2 \leq\frac{1}{D}\cdot \min_{d\in\mathcal{D}}  I_d(\overline{X}_1,\overline{X}_2;\overline{Y}|Q)\\
&\hspace{1cm}R_2 \leq   \frac{1}{D}\cdot \min_{d\in\mathcal{D}}  I_d(\overline{X}_2;\overline{Y}|\tilde{V},Q)
\end{array}\right\}.
\end{flalign}
where $\tilde{V}$ is distributed according to (\ref{eq:v_bar_vec_outer}), and $P_d(q,\overline{x}_1,\tilde{v},\overline{x}_2,\overline{y})$ is defined according to (\ref{eq:outer_prob}).
Then, the closure  of $\overline{\mathcal{R}}$ includes the achievable region of  the ACMAC.
\end{theorem}
The proof of this theorem appears in Appendix~\ref{ACMAC_C2}.
It should be noted that the random variables appearing in (\ref{outer_region_ACMAC}), $\bar{X}_1, \bar{X_2}, \bar{Y}, \tilde{V}$, take values in the alphabets $\mathcal{X}_1^D,\mathcal{X}_2^D,\mathcal{Y}^D,\mathcal{X}_1^{2D-1}$, respectively. Therefore, (\ref{outer_region_ACMAC}) is in fact a single-letter expression in the sense that the alphabet cardinalities involved do no increase with the blocklength $n$.

Further, one can consider a different setup in which the delay $d$ symbolizes the presence of a jitter. The jitter is modeled by a delay that randomly changes every sub-block of a sufficiently large size that allows the decoder to find the delay in the sub-block with an error probability that decays as the block length tends to infinity. It can be shown that in this setup, if the delays are i.i.d. random variables distributed over the set $\mathcal{D}$ and if each sub-block length tends to infinity with the codeword length, then the minimizations over the delay $d$ in Theorems~\ref{ACMAC_inner_single} and \ref{ACMAC_outer_single} (in Equations (\ref{inner_region_ACMAC}) and (\ref{outer_region_ACMAC})) can be replaced with an expectation over the delay $d$.

\begin{remark}
Consider the binary ACMAC defined by the following inputs-output relation,
\begin{flalign}\label{Binary_ACMAC_setup}
Y_i=X_{1,i-d}\oplus X_{2,i}\oplus Z_i
\end{flalign}
where $Z_i \sim \text{Bernoulli}\left(p\right)$, and  $d\in\mathcal{D}$.
Assume that the informed encoder (encoder 2) knows in advance the message of the non-cognitive user, which can be regarded as a common message.

We note that under the synchronous case in which $d_{min}=d_{max}=0$ the outer and inner regions (Theorems~\ref{ACMAC_inner_single} and \ref{ACMAC_outer_single}) coincide and the resulting expression is the capacity region of the CMAC. It is easy to verify that in the binary setup (\ref{Binary_ACMAC_setup}), the capacity regions of the synchronous cognitive and non-cognitive MACs  are equal, that is, the side information does not enlarge the capacity region. Consequently, the capacity regions of the binary ACMAC and binary MAC are equal. That is, the capacity region of the binary ACMAC is the union of all rate-pairs $(R_1,R_2)$ such that
$R_1+R_2\leq 1- H(Z)$.
\end{remark}
\section{The Gaussian ACMAC with Individual Power Constraints}\label{sec:Examples_ACMAC}
Consider the Gaussian ACMAC defined by the inputs-output relation,
\begin{flalign}\label{Gassian_ACMAC_model}
Y_i=X_{1,i-d} + X_{2,i} + Z_i
\end{flalign}
where $Z_i \sim \mathcal{N}(0,N)$, and  each transmitter obeys an individual power constraint,
\begin{flalign}\label{power_const_n}
&\frac{1}{n}\sum_{i=1}^nX_{1,i}^2\leq P_1,\quad\frac{1}{n}\sum_{i=1}^nX_{2,i}^2\leq P_2.
\end{flalign}
For clarity of the presentation we consider
\begin{flalign}\label{eq:set_d_Gauss}
 d\in\mathcal{D}=\{0,1\}.
 \end{flalign}
It is assumed that the informed encoder knows in advance the message of the non-cognitive user (common message).
We emphasize that in the case, we consider a variant of the Gaussian ACMAC  in which the symbols of the uninformed encoder are statistically independent. In this scenario the first encoder does not wish to deviate from  its simple independently generated  codebook in order to possibly achieve higher rates. We refer to this setup by the term  \textit{sub-cooperative  Gaussian ACMAC}.

We next present outer and inner bounds on the capacity region of the Gaussian ACMAC.

\textbf{The Outer Bound:}
In the following we discuss the outer bound of the Gaussian ACMAC and the sub-cooperative ACMAC.
\begin{lemma}\label{lemma:lemma_Gaussian}
Under the Gaussian channel model with individual power constraints which is given in (\ref{Gassian_ACMAC_model})-(\ref{eq:set_d_Gauss}), it is sufficient to take the union over a deterministic $Q$ and jointly Gaussian $\tilde{V}$ and $\overline{X}_{2}$
in the rate-region (\ref{outer_region_ACMAC}). Additionally, it is  sufficient to consider random vectors $\tilde{V}$ and $\overline{X}_{2}$
 such that\footnote{\label{foot:expec_d}The expression $E_d\left[\overline{X}_{1}\overline{X}_{1}^T\right]$ is the expected value of $\overline{X}_{1}\overline{X}_{1}^T$ for a delay $d\in\mathcal{D}$.}
\begin{flalign}\label{power_constraints_text}
E_d\left[\overline{X}_{1}\overline{X}_{1}^T\right]\leq 2P_1\quad , E\left[\overline{X}_{2}\overline{X}_{2}^{T}\right]\leq2P_2,\quad \forall d\in\mathcal{D}
\end{flalign}
where $\overline{X}_1$ and $\overline{X}_2$ are defined as (\ref{eq:dep_over_X1}) and (\ref{eq:dep_over_X2}), respectively.
\end{lemma}

The proof of this lemma appears in Appendix~\ref{ACMAC_Gaussian1}. The proof is established by Lemma 1 in \cite{Thomas1987} which claims that for any two random vector $W_{1}$ and $W_{2}$ such that the random vector $(W_{1},W_{2})$ has a covariance matrix $C$ the conditional entropy $h(W_{1}|W_{2})$ is maximized by  jointly Gaussian $W_{1}$ and $W_{2}$.

\begin{proposition}\label{props:Gauss_ACMAC_sub_coop}
The region
\begin{flalign}\label{outer_Gauss_region_ACMAC_ind}
\bigcup_{\rho\in\left[0,\frac{1}{\sqrt{2}}\right]}
\left\{
\begin{array}{ll}
    \multirow{2}{*}{\hspace{-0.2cm}$(R_1,R_2):$}\hspace{-0.3cm}&R_1+R_2 \leq\frac{1}{2}\log \left(1+\frac{P_{1}+2\rho\sqrt{P_1P_2}+P_2}{N}\right)\\
&\hspace{1cm}R_2 \leq   \frac{1}{2}\log\left(1+\frac{P_{2}(1-2\rho^2)}{N}\right)
\end{array}\right\}.
\end{flalign}
is an outer region for the sub-cooperative  Gaussian ACMAC with $\mathcal D=\{0,1\}$.
\end{proposition}
The proof of this lemma appears in Appendix~\ref{ACMAC_Gaussian2}.

\textbf{The Inner Bound:}
The following theorem states an inner bound on the capacity region of the Gaussian ACMAC.
\begin{proposition}\label{proposition_Gaussian_inner}
The  rate-pairs $(R_1,R_2)$ satisfying
\begin{flalign}
&R_1+R_2\leq
\frac{1}{2}\log\left(\frac{N+P_1+\tilde{P}_2+2\rho\sqrt{P_1\tilde{P}_2}}
{N+\tilde{P}_2(1-\rho^2)}
\right)+
\frac{1}{2}\log\left(1+\frac{\tilde{P}_2}{N}(1-2\rho^2)\right)
\nonumber\\
&R_2\leq \frac{1}{2}\log\left(1+\frac{\tilde{P}_2}{N}(1-2\rho^2)\right)
\end{flalign}
for some $\tilde{P}_2\in[0,P_2]$ and $\rho\in \left[0,\frac{ 1}{\sqrt{2}}\right]$, is an inner bound on the capacity of the Gaussian ACMAC with $\mathcal{D}=\{0,1\}$.
\end{proposition}

\begin{IEEEproof}

We will deduce an inner bound on the region (\ref{inner_region_ACMAC}), by choosing $(\textbf{V},X_2)$ which are jointly Gaussian.

Let $d\in\{0,1\}$ be a given delay and let $\overline{V}=(V_1,V_2)$ where $X_1=V_1\cdot\mathbbm{1}_{\{d=1\}}+V_2\cdot\mathbbm{1}_{\{d=0\}}$.
Further, let $(V_1,V_2,X_2)$ be jointly Gaussian with zero mean, and the following covariance matrix\footnote{We remark that this choice of covariance matrix fulfills the power constraints with probability that tends to one as $n$ tends to infinity (by the LLN).}:
\begin{flalign}\label{covariance_inner}
C_{V_1,V_2,X_2}=
\begin{pmatrix}
\tilde{P}_1 & 0 & \sigma_1\\
0 & \tilde{P}_1 & \sigma_2\\
\sigma_1 & \sigma_2 &\tilde{P}_2
\end{pmatrix},
\end{flalign}
where $\tilde{P}_1\leq P_1$, and $\tilde{P}_2\leq P_2$ (see (\ref{power_constraints_text})).

It is easy to verify that for this choice of random variables one has
\begin{flalign}
I_d(X_2;Y|\overline{V})=\frac{1}{2}\log\left(1+\frac{\tilde{P}_1\tilde{P}_2
-(\sigma_1^2+\sigma_2^2)}{N\tilde{P}_1}\right).
\end{flalign}
We now consider $I_d(X_1;Y)$
\begin{flalign}
I_d(X_1;Y)&=h_d(Y)-h_d(Y|X_1)\nonumber\\
&=h_d(X_1+X_2+Z)-h_d(X_1+X_2+Z|X_1)\nonumber\\
&=h_d(X_1+X_2+Z)-h_d(X_2+Z|X_1)\nonumber\\
&\stackrel{(a)}{=}h_d(X_1+X_2+Z)-h_d(X_2+Z-E[X_2+Z|X_1]|X_1)\nonumber\\
&\stackrel{(b)}{=}h_d(X_1+X_2+Z)-h_d(X_2+Z-E[X_2+Z|X_1])\nonumber\\
&=h_d(X_1+X_2+Z)-h_d(X_2+Z-E[X_2|X_1])
\end{flalign}
where (a) follows since $E[X_2+Z|X_1]$ is a function of $X_1$, and
(b) follows since $E[(X_2+Z-E[X_2+Z|X_1])X_1]=0$,
$X_1=V_1\cdot\mathbbm{1}_{\{d=1\}}+V_2\cdot\mathbbm{1}_{\{d=0\}}$, and since $(V_1,V_2,X_2,Z)$ are jointly Gaussian.

Now, let $\Lambda_d(\sigma_1,\sigma_2)=\sigma_1\cdot\mathbbm{1}_{\{d=1\}}+\sigma_2\cdot\mathbbm{1}_{\{d=0\}}$.
Given the delay $d$, $X_1$ is either $V_1$ or $V_2$, also each of the vector $(V_1,X_2,Z)$ and $(V_1,X_2,Z)$ is jointly Gaussian, it follows that
$E[X_2|X_1]=\Lambda_d(\sigma_1,\sigma_2)\tilde{P}_1^{-1}X_1$. Therefore,
\begin{flalign}
h_d(X_2+Z-E[X_2|X_1]) &= h_d(X_2+Z-\Lambda_d(\sigma_1,\sigma_2)\tilde{P}_1^{-1}X_1) \nonumber\\
&= \frac{1}{2}\log(2\pi e)+\frac{1}{2}\log\left(N+\tilde{P}_2+\frac{\Lambda_d(\sigma_1,\sigma_2)^2}{\tilde{P}_1}-2\frac{\Lambda_d(\sigma_1,\sigma_2)^2}{\tilde{P}_1}\right).
\end{flalign}
In addition,
\begin{flalign}
h_d(X_1+X_2+Z)=\frac{1}{2}\log(2\pi e)+\frac{1}{2}\log\left(N+\tilde{P}_1+\tilde{P}_2+2\Lambda_d(\sigma_1,\sigma_2)\right).
\end{flalign}
Therefore,
\begin{flalign}
&I_d(X_1;Y)=\frac{1}{2}\log\left(\frac{N+\tilde{P}_1+\tilde{P}_2+2\Lambda_d(\sigma_1,\sigma_2)}
{N+\tilde{P}_2-\frac{\Lambda_d(\sigma_1,\sigma_2)^2}{\tilde{P}_1}}
\right)\nonumber\\
&I_d(X_2;Y|\overline{V})=\frac{1}{2}\log\left(1+\frac{\tilde{P}_1\tilde{P}_2
-(\sigma_1^2+\sigma_2^2)}{N\tilde{P}_1}\right).
\end{flalign}
Now, let $\rho_1=\frac{\sigma_1}{\sqrt{\tilde{P_1}{\tilde{P_2}}}}$, and $\rho_2=\frac{\sigma_2}{\sqrt{\tilde{P_1}{\tilde{P_2}}}}$, it follows that
\begin{flalign}
&I_d(X_1;Y)=\frac{1}{2}\log\left(\frac{N+\tilde{P}_1+\tilde{P}_2+2\sqrt{\tilde{P}_1\tilde{P}_2}\Lambda_d(\rho_1,\rho_2)}
{N+\tilde{P}_2[1-\Lambda_d(\rho_1,\rho_2)^2]}
\right)\nonumber\\
&I_d(X_2;Y|\overline{V})=\frac{1}{2}\log\left(1+\frac{\tilde{P}_2[1-(\rho_1^2+\rho_2^2)]}{N}\right).
\end{flalign}

Hence, the following region, denoted  $\underline{\mathcal{R}}(\tilde{P}_1,\tilde{P}_2,\rho_1,\rho_2)$
 \begin{flalign}
\left\{
\begin{array}{ll}
\multirow{2}{*}{$(R_1,R_2):$}
&R_1+R_2 \leq \min_{d\in\{0,1\}}\left[
\frac{1}{2}\log\left(\frac{N+\tilde{P}_1+\tilde{P}_2+2\sqrt{\tilde{P}_1\tilde{P}_2}\Lambda_d(\rho_1,\rho_2)}
{N+\tilde{P}_2[1-\Lambda_d(\rho_1,\rho_2)^2]}\right)\right. \nonumber\\
&\hspace{6cm}\left. \vphantom{\frac{1}{2}\log\left(\frac{N+\tilde{P}_1+\tilde{P}_2+2\sqrt{\tilde{P}_1\tilde{P}_2}\Lambda_d(\rho_1,\rho_2)}
{N+\tilde{P}_2[1-\Lambda_d(\rho_1,\rho_2)^2]}\right)}+\frac{1}{2}\log\left(1+\frac{\tilde{P}_2[1-(\rho_1^2+\rho_2^2)]}{N}\right)
\right]\nonumber\\
&R_2 \leq \frac{1}{2}\log\left(1+\frac{\tilde{P}_2}{N}(1-(\rho_1^2+\rho_2^2))\right)
\end{array}
\right\}.
\end{flalign}\
is achievable.

Now, one can pick\footnote{It can be shown,  that this choice does not reduce the convex-hull of all the rate-pairs $(R_1,R_2)$ such that $(R_1,R_2)\in\underline{\mathcal{R}}(\tilde{P}_1,\tilde{P}_2,\rho_1,\rho_2)$,
for some $\rho_1,\rho_2\in[-1,1]$ such that $\rho_1^2+\rho_2^2\leq1$, $\tilde{P}_1\in[0,P_1]$, and $\tilde{P}_2\in[0,P_2]$.}
 $\tilde{P}_1=P_1$, and $\rho_1=\rho_2\in \left[0,\frac{1}{\sqrt{2}}\right]$
, and consequently, the convex hull of all rate-pairs $(R_1, R_2)$ satisfying
\begin{flalign}
&R_1+R_2\leq
\frac{1}{2}\log\left(\frac{N+P_1+\tilde{P}_2+2\rho\sqrt{P_1\tilde{P}_2}}
{N+\tilde{P}_2(1-\rho^2)}
\right)+
\frac{1}{2}\log\left(1+\frac{\tilde{P}_2}{N}(1-2\rho^2)\right)
\nonumber\\
&R_2\leq \frac{1}{2}\log\left(1+\frac{\tilde{P}_2}{N}(1-2\rho^2)\right)
\end{flalign}
for some $\tilde{P}_2\in[0,P_2]$ and $\rho\in \left[0,\frac{ 1}{\sqrt{2}}\right]$, is an inner bound on the capacity of the Gaussian ACMAC.
\end{IEEEproof}
\begin{figure}[H]
\centering
\includegraphics[scale=0.55]{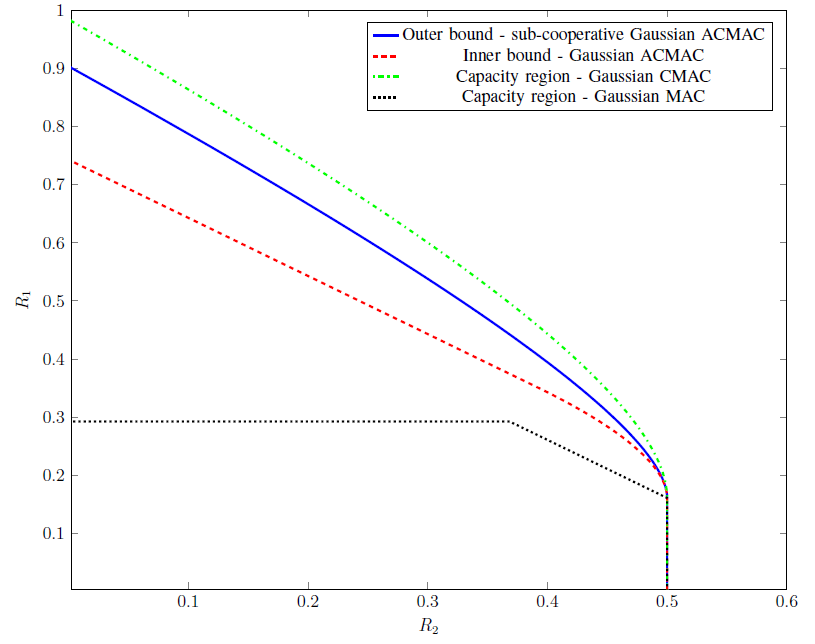}\\
\caption{Outer and inner bounds on the capacity region of the Gaussian ACMAC/sub-cooperative Gaussian ACMAC with $d\in\{0,1\}$ compared with the capacity region of the Gaussian CMAC with parameters $P_1=0.5,P_2=N=1$.}
\label{Gaussian_ACMAC_fig}
\end{figure}
Fig.\ \ref{Gaussian_ACMAC_fig} compares the capacity region of the synchronous Gaussian CMAC, whose capacity is derived in \cite[Theorem 7]{Anelia2008}, and the outer and inner bounds on the capacity regions of the Gaussian ACMAC/sub-cooperative Gaussian ACMAC  with $d\in\{0,1\}$, $P_1=0.5$, and $P_2=N=1$.

\section{Bounds on the Capacity Region of the ACC-MAC}\label{ACC-MAC_sec}

In this section we derive the capacity region of the ACC-MAC in terms of a multi-letter expression and state single-letter  inner and outer bounds on it.
We prove that the capacity region of the ACC-MAC is included in the capacity region of the respective ACMAC.
This is due to the fact that when message cognition is concerned, rate-splitting techniques can be adopted, that is, the uninformed encoder can transmit only part of its message knowing that the cognitive encoder can transmit the rest. However, when codeword side-information is considered the informed encoder can only find the uninformed encoder's message by decoding its codeword. For this reason, unlike the ACMAC model, the informed encoder cannot help the uninformed encoder achieve rates which exceed its codeword's entropy.

\subsection{A Multi-letter Expression for the Capacity Region of the ACC-MAC}

Recall the definition of $P_d(\textbf{y}|\textbf{x}_1,\textbf{x}_2)$  (see (\ref{trans_prob_ACMAC})), and let
 $P_d(\textbf{x}_1,\textbf{x}_2,\textbf{y})=P(\textbf{x}_1,\textbf{x}_2)P_d(\textbf{y}|\textbf{x}_1,\textbf{x}_2)$. 

Define the region $\mathcal{R}_n$ of rate-pairs $(R_1,R_2)$:
\begin{flalign}\label{R_n_def_ACCMAC}
&\check{\mathcal{R}}_n=\bigcup_{P(\textbf{x}_1,\textbf{x}_2)}\bigcap_{d\in\mathcal{D}}
\left\{(R_1,R_2):
\begin{array}{ll}
    \multirow{2}{*}{} &\hspace{1cm}R_1 \leq \frac{1}{n} H(\textbf{X}_1),\\
    &\hspace{1cm}R_2 \leq \frac{1}{n}I_d(\textbf{X}_2;\textbf{Y}|\textbf{X}_1),\\
    & R_1 +R_2 \leq \frac{1}{n}I_d(\textbf{X}_1,\textbf{X}_2;\textbf{Y})
\end{array}
\right\},
\end{flalign}
and the region $\mathcal{Q}_n$ of rate-pairs $(R_1,R_2)$:
\begin{flalign}
&\check{\mathcal{Q}}_n=\bigcup_{P(\textbf{x}_1,\textbf{x}_2)}\bigcap_{d\in\mathcal{D}}
\left\{(R_1,R_2):
\begin{array}{ll}
    \multirow{2}{*}{}
   & \hspace{1cm}R_1 \leq \frac{1}{n}H(\textbf{X}_1),\\
    &\hspace{1cm}R_2 \leq \frac{1}{n}I_{d}(\textbf{X}_2;Y_{d_{max}+1}^{n-d_{min}}|\textbf{X}_1),\\
    & R_1 +R_2 \leq \frac{1}{n}I_{d}(\textbf{X}_1,\textbf{X}_2;Y_{d_{max}+1}^{n-d_{min}})
\end{array}
\right\}
\end{flalign}
where
$P_{d}(y_{d_{max}+1}^{n-d_{min}}|\textbf{x}_1,\textbf{x}_2)=\prod_{i=d_{max}+1}^{n-d_{min}}P(y_i|x_{1,i-d},x_{2,i})$.

We can now state the capacity region of the ACC-MAC in terms of a multi-letter expression.
\begin{theorem}\label{ACC-MAC_capacity}
Let $P_{Y|X_1,X_2}$ be an ACC-MAC with a finite set of possible delays $\mathcal{D}$. The capacity region of the ACC-MAC is given by:
\begin{flalign}\label{ACC-MAC_capacity_term}
\mathcal{C}&=\text{closure}\left(\bigcup_{n \geq D} \check{\mathcal{Q}}_n\right)=
\text{closure}(\lim\sup_{n\rightarrow\infty}\check{\mathcal{R}}_n)=\text{closure}(\lim\inf_{n\rightarrow\infty}\check{\mathcal{R}}_n).
\end{flalign}
\end{theorem}
The outline of the proof of achievability part of Theorem~\ref{ACC-MAC_capacity} appears in  Appendix~\ref{ACCMAC_C1}.
The additional inequality compared to the multi-letter expression of the capacity region of the ACMAC, Eq. (\ref{ACMAC_capacity_term}), $R_1 \leq \frac{1}{n}H(\textbf{X}_1)$ follows since
\begin{flalign}
nR_1= H(M_1)=H(\textbf{X}_1,M_1)=H(\textbf{X}_1)+H(M|\textbf{X}_1),
\end{flalign}
where $\frac{1}{n}H(M|\textbf{X}_1)\rightarrow0$ as $n\rightarrow \infty$ by Fano's inequality and since the informed encoder possesses only codeword side-information.

Comparing (\ref{ACMAC_capacity_term}) and (\ref{ACC-MAC_capacity_term}), it is easy to see that, as expected, $\mathcal{C}_{ACC-MAC}\subseteq \mathcal{C}_{ACMAC}$ where $\mathcal{C}_{ACMAC}$ and $\mathcal{C}_{ACC-MAC}$ are the capacity regions of the ACMAC and ACC-MAC, respectively.

\subsection{Inner and Outer Bounds on the Capacity Region of the ACC-MAC }
We proceed to present  inner and outer bounds on the capacity region of the ACC-MAC.
In this section we use the notations of Section~\ref{ACMAC_bounds}.

\begin{theorem}\label{ACC-MAC_inner_single}
Let $P_{Y|X_1,X_2}$ be an ACC-MAC and let $(X_1,\overline{V},X_2,Y)$ be distributed according to (\ref{acmac_single_def}).
Denote,
\begin{flalign}\label{inner_region_ACC-MAC}
&\underline{\mathcal{R}}=\bigcup_{P(x_1),P(x_2|\overline{v})}\left\{
(R_1,R_2):\begin{array}{ll}
    \multirow{2}{*}{}&\hspace{1cm}R_1\leq H(X_1),\\
    &\hspace{1cm}R_2 \leq \min_{d\in\mathcal{D}}I_d(X_2;Y|\overline{V}),\\
    & R_1 +R_2 \leq \min_{d\in\mathcal{D}}\left[I_d(X_1;Y)+I_d(X_2;Y|\overline{V})\right]
\end{array}
\right\}.
\end{flalign}
The closure convex  of $\underline{\mathcal{R}}$ is an achievable rate region for the ACC-MAC.
\end{theorem}

We note that the coding scheme of Theorem~\ref{ACMAC_inner_single} is not suited to this theorem. This is due to the fact that the informed encoder knows the uninformed encoder's message by decoding the uninformed encoder's codeword. This leads to the conclusion that in the ACC-MAC model the mapping $f(m_1)$ (see Section \ref{sec:Channel Models}) must be reversible,  unlike the ACMAC model for which we need not make such a stipulation. In this case  simultaneous decoding yields better results compared with successive decoding.

The proof of Theorem~\ref{ACC-MAC_inner_single} appears in Appendix~\ref{ACCMAC_A2}.

As in the ACMAC model, we can improve the result of Theorem~\ref{ACC-MAC_inner_single} by minimizing the expected terms in (\ref{inner_region_ACC-MAC}). We state it formally in the following corollary.
\begin{corollary}
Let $P_{Y|X_1,X_2}$ be an ACC-MAC with a finite set of possible delays $\mathcal{D}$.
Let $(Q,X_1,\overline{V},X_2,Y)$ be distributed according to  (\ref{acmac_single_def_q}).
Denote,
\begin{flalign}\label{inner_region_ACMAC_q}
\underline{\mathcal{R}}&=\bigcup_{P(q),P(x_1|q),P(x_2|\overline{v},q)}
\left\{
\begin{array}{ll}
    \multirow{2}{*}{$(R_1,R_2):$} &\hspace{1cm}R_1\leq H(X_1|Q),\\
    &R_1 +R_2 \leq \min_{d\in\mathcal{D}}\left[I_d(X_1;Y|Q)+I_d(X_2;Y|\overline{V},Q)\right],\\
    &\hspace{1cm}R_2 \leq \min_{d\in\mathcal{D}}I_d(X_2;Y|\overline{V},Q)
\end{array}
\right\}.
\end{flalign}
The  closure   of $\underline{\mathcal{R}}$ is an achievable rate region for the ACMAC.
\end{corollary}

We next provide an outer bound on the capacity region of the ACC-MAC.
\begin{theorem}\label{ACC-MAC_outer_single}
Let $P_{Y|X_1,X_2}$ be an ACC-MAC with a finite set of possible delays $\mathcal{D}$. Denote,
\begin{flalign}\label{outer_region_ACC-MAC}
\overline{\mathcal{R}}&=\bigcup_{P(q),P(\tilde{v}|q),P(\overline{x}_2|\tilde{v},q)}
\left\{
\begin{array}{ll}
    \multirow{2}{*}{$(R_1,R_2):$}&\hspace{1cm}R_1\leq \frac{1}{D}\cdot\min_{d\in\mathcal{D}}  H_{d}(\overline{X}_1|Q)\\
    &R_1+R_2 \leq\frac{1}{D}\cdot \min_{d\in\mathcal{D}}  I_d(\overline{X}_1,\overline{X}_2;\overline{Y}|Q)\\
&\hspace{1cm}R_2 \leq   \frac{1}{D}\cdot \min_{d\in\mathcal{D}} I_d(\overline{X}_2,\overline{Y}|\tilde{V},Q)
\end{array}\right\}.
\end{flalign}
where $\tilde{V}$ is distributed according to (\ref{eq:v_bar_vec_outer}), and $P_d(q,\overline{x}_1,\tilde{v},\overline{x}_2,\overline{y})$ is defined according to (\ref{eq:outer_prob}).
Then, the closure  of $\overline{\mathcal{R}}$ includes the achievable region of  the ACC-MAC.
\end{theorem}
The proof of this theorem is similar to the proof of Theorem~\ref{ACMAC_outer_single} and therefore it is omitted and we only outline the differences between the proofs.  The additional inequality $R_1\leq \frac{1}{D}\cdot\min_{d\in\mathcal{D}}  H_{d}(\overline{X}_1|Q)$ is obtained by the following argument,
\begin{flalign}
nR_1&= H(M_1)\stackrel{(*)}{=}\min_{d\in\mathcal{D}}H_{d}(\textbf{X}_1,M_1)=\min_{d\in\mathcal{D}}[H_{d}(\textbf{X}_1)+H_{d}(M|\textbf{X}_1)].
\end{flalign}
The equality $(*)$ follows since for every $d\in\mathcal{D}$, the codeword $\textbf{X}_1$ that is transmitted over the channel is a deterministic function of the message $M_1$ and the delay.

As in the proof of Theorem~\ref{ACMAC_outer_single} (see the line before Eq. (\ref{MAC_upper_1})), denote $\overline{X}_{1,i}=X_{1,i D+1-d}^{(i+1) D-d}$. Further, let $Q$ be a random variable that is distributed uniformly over the set $\{0,\ldots,n/D-1\}$ and let $\overline{X}_{1}\triangleq\overline{X}_{1,Q}$. Then
\begin{flalign}
\frac{1}{n} H_{d}(\textbf{X}_1)\leq
\frac{1}{n}\sum_{i=0}^{n/D-1}H_d(\overline{X}_{1,i})=\frac{1}{D}\cdot
\frac{D}{n}\sum_{i=0}^{n/D-1}H_d(\overline{X}_{1,i})=\frac{1}{D}H_d(\overline{X}_{1}|Q).
\end{flalign}
Note that as before $\frac{1}{n}H_{d}(M|\textbf{X}_1)\rightarrow0$ as $n\rightarrow \infty$ by Fano's inequality and since the informed encoder possesses only codeword side-information.

We further remark that the regions (\ref{ACC-MAC_capacity_term}), (\ref{inner_region_ACC-MAC}) and (\ref{outer_region_ACC-MAC}) differ from the regions of Theorem~\ref{ACMAC_capacity}, \ref{ACMAC_inner_single} and \ref{ACMAC_outer_single}, respectively, in the additional inequality on the rate of the uniformed user that is added in the ACC-MAC model.

Consider the Gaussian ACC-MAC with the same characteristics as those of the ACMAC which is presented in Section~\ref{sec:Examples_ACMAC}. The informed encoder receives the codeword of the uninformed one without any additional noise. Further,  for each $\epsilon>0$ the open interval  $(-\epsilon,\epsilon)$ includes an uncountable number of real numbers. Combining these two facts, we  infer that the uninformed encoder can use its first symbol in each codeword to notify the cognitive user about its message while consuming a negligible amount of power. Thus, the capacity regions of the Gaussian ACC-MAC and the Gaussian ACMAC (Section~\ref{sec:Examples_ACMAC}) coincide. Further, the outer and inner bounds on the capacity region of the Gaussian ACMAC, i.e., Lemma~\ref{lemma:lemma_Gaussian} and Propositions~\ref{props:Gauss_ACMAC_sub_coop} and~\ref{proposition_Gaussian_inner}, hold for the Gaussian ACC-MAC.

\section{An Example - Comparison between the ACC-MAC and the ACMAC Models}\label{sec:example}

Let $\mathcal{X}_1=\{2,4\}$, and let $\mathcal{X}_2=\mathcal{Y}=\{0,1,2,3\}$. Define the channel by the following inputs-output relation:
\begin{flalign}\label{example:channel_model_ACMAC}
Y= X_2\pmod{X_1}.
\end{flalign}

We first analyze the case of no delay ($\mathcal{D}=\{0\}$) as an example which demonstrates that the capacity region of the ACC-MAC can be strictly smaller than that of the ACMAC.
Let CMAC and CC-MAC denote the synchronous setups of the ACMAC and ACC-MAC models, respectively.
Since the channel is synchronous, we can use the results of \cite{SlepianWolf1973} to establish that the capacity region of the CMAC is given by:
\begin{flalign}
&\mathcal{C}_{CMAC}=\bigcup_{P(q),P(x_1,x_2|q)}\left\{(R_1,R_2):\begin{array}{ll}
\multirow{2}{*}{}
R_2 &\leq I(X_2;Y|X_1,Q),\\
R_1+R_2 &\leq I(X_1,X_2;Y|Q)
\end{array}
\right\}
\end{flalign}
It follows that:
\begin{flalign}
&I(X_1,X_2;Y|Q)\leq\ H(Y)\leq \log |\mathcal{Y}|=2\text{ bits}\nonumber\\
&I(X_2;Y|X_{1},Q)\leq H(Y|X_1)\leq \log |\mathcal{Y}|=2 \text{ bits},
\end{flalign}
with equalities if $\Pr(x_1=4)=1$, and $\Pr(x_2)=\frac{1}{4}$ for all $x_2\in\mathcal{X}_2$.
Consequently, the capacity of the proposed channel is the triangle:
 \begin{flalign} \label{example:channel_model_ACMAC_capacity_region}
 &\mathcal{C}_{CMAC}=\left\{(R_1,R_2):\begin{array}{ll}
 \multirow{2}{*}{}
 R_1+R_2 &\leq 2
 \end{array}
 \right\}.
 \end{flalign}

A coding scheme that achieves every rate-pair in the capacity region lets the informed encoder send the messages of both encoders. Therefore, the capacity region of the channel given in (\ref{example:channel_model_ACMAC}) under the CMAC model is unaffected by asynchronism regardless of the delay $\mathcal{D}$; it follows that $\mathcal{C}_{ACMAC}=\mathcal{C}_{CMAC}$.

Under the CC-MAC model (ACC-MAC with $\mathcal{D}=\{0\}$), the inner and outer bounds, i.e., the regions (\ref{inner_region_ACC-MAC}) and (\ref{outer_region_ACC-MAC}), respectively, coincide with the capacity region:
\begin{flalign}\label{example:channel_model_ACC-MAC}
&\mathcal{C}_{CC-MAC}=\bigcup_{P(q),P(x_1,x_2|q)}\left\{(R_1,R_2):\begin{array}{ll}
\multirow{2}{*}{}
R_1 &\leq H(X_1|Q),\\
R_2 &\leq I(X_2;Y|X_1,Q),\\
R_1+R_2 &\leq I(X_1,X_2;Y|Q)
\end{array}
\right\}.
\end{flalign}
Now,
 \begin{flalign}
 I(X_2;Y|X_1,Q)&=H(Y|X_1,Q)\nonumber\\&\leq H(Y|X_1)\nonumber\\&=\Pr(x_1=4)H(X_2)+\Pr(x_1=2)H(X_2\pmod{2}).
 \end{flalign}
 It can be shown that the capacity region is the following trapezoid
 \begin{flalign}\label{example_ACC-MAC_cpacity_region}
&\mathcal{C}_{CC-MAC}=\left\{(R_1,R_2):\begin{array}{ll}
\multirow{2}{*}{}
R_1 &\leq 1,\\
R_1+R_2 &\leq 2
\end{array}
\right\}.
\end{flalign}
Its two corner points are achieved by the p.m.f.'s $\Pr(x_1=4)=\Pr(x_1=2)=\frac{1}{2},\Pr(x_2=2)=\Pr(x_2=3)=\frac{1}{2}$, and
 $\Pr(x_1=4)=1,\Pr(x_2)=\frac{1}{4}\text{ } \forall x_2\in\mathcal{X}_2$.
The region (\ref{example_ACC-MAC_cpacity_region}) is equivalent to the capacity region of the channel given in (\ref{example:channel_model_ACMAC}) with no side-information at both transmitters.
Therefore, under CC-MAC, the side-information in the this channel does not enlarge the capacity region, and we obtain $\mathcal{C}_{CC-MAC} \subset \mathcal{C}_{CMAC}$.

Moreover, since the region  (\ref{example_ACC-MAC_cpacity_region}) can be achieved by a coding scheme that does not use the side-information, we can use \cite{CoverMcEliece1981} to deduce that bounded asynchronism does not affect the channel capacity region.
Consequently, $\mathcal{C}_{ACC-MAC} \subset \mathcal{C}_{ACMAC}$ for every set $\mathcal{D}$ of bounded delays.

\section{Conclusion}\label{sec:Conclusion}
 In this paper we presented the asynchronous cognitive MAC with message and/or codeword cognition at one encoder, denoted ACMAC and ACC-MAC, respectively. We characterized the capacity regions of the ACMAC and ACC-MAC in terms of multi-letter expressions. We presented  inner and outer bounds on the capacity regions of these channels. Further, we analyzed the Gaussian ACMAC and derived inner and outer bounds on it capacity region.
   We noted that in the Gaussian case the capacity regions of the ACMAC and the ACC-MAC are equal.
   Finally, we presented an example for a channel in which the ACC-MAC capacity region is strictly smaller than the capacity region of the ACMAC.
\appendix
\subsection{An auxiliary lemma}\label{lemma1}

The following lemma will be used in the proof of Theorem~\ref{ACMAC_capacity} (see Appendix \ref{ACMAC_C1}).
\begin{lemma}\label{lemma_discard}
Let $\textbf{X},\textbf{Y}$ and $\textbf{Z}$ be random vectors whose symbols belong to the finite alphabets $\mathcal{X},\mathcal{Y}$ and $\mathcal{Z}$, respectively.
Additionally, let $\Theta$ be a finite set, let $\theta\in\Theta$, and denote
\begin{flalign}\label{lemma1_begin}
&P_{\theta,\textbf{X},\textbf{Y},\textbf{Z}}=P_{\textbf{X},\textbf{Y}}P_{\theta,\textbf{Z}|\textbf{X},\textbf{Y}}\nonumber\\
&Q_{\theta,\textbf{X},\textbf{Y},\textbf{Z}}=P_{\textbf{X},\textbf{Y}}Q_{\theta,\textbf{Z}|\textbf{X},\textbf{Y}}
\end{flalign}
where
\begin{flalign}
&P_{\theta,\textbf{Z}|\textbf{X},\textbf{Y}}=P_{\theta,Z_1^{D_1}|\textbf{X},\textbf{Y}}P_{\theta,Z_{n-D_2+1}^n|\textbf{X},\textbf{Y}}P_{\theta,Z_{D_1+1}^{n-D_2}|\textbf{X},\textbf{Y}}\nonumber\\
&Q_{\theta,\textbf{Z}|\textbf{X},\textbf{Y}}=Q_{\theta,Z_1^{D_1}|\textbf{X},\textbf{Y}}Q_{\theta,Z_{n-D_2+1}^n|\textbf{X},\textbf{Y}}P_{\theta,Z_{D_1+1}^{n-D_2}|\textbf{X},\textbf{Y}}
\end{flalign}
and $D_1,D_2$ are nonnegative finite integers.

Denote by $\mathcal{R}_{P,n}$ and $\mathcal{R}_{Q,n}$ the following regions
\begin{flalign}
&\mathcal{R}_{P,n}=\bigcup_{P(\textbf{x},\textbf{y})}\bigcap_{\theta\in\Theta}{}\left\{
\begin{array}{ll}
    \multirow{2}{*}{$(R_1,R_2):$}&R_1+R_2 \leq \frac{1}{n}I_{P_{\theta}}(\textbf{X},\textbf{Y};\textbf{Z})\\
&\hspace{1cm}R_2 \leq \frac{1}{n}I_{P_{\theta}}(\textbf{Y};\textbf{Z}|\textbf{X})
\end{array}
\right\}\nonumber\\
&\mathcal{R}_{Q,n}=\bigcup_{P(\textbf{x},\textbf{y})}\bigcap_{\theta\in\Theta}\left\{
\begin{array}{ll}
    \multirow{2}{*}{$(R_1,R_2):$}&R_1+R_2 \leq \frac{1}{n}I_{Q_{\theta}}(\textbf{X},\textbf{Y};\textbf{Z})\\
&\hspace{1cm}R_2 \leq \frac{1}{n}I_{Q_{\theta}}(\textbf{Y};\textbf{Z}|\textbf{X})
\end{array}
\right\}.
\end{flalign}

Then,
\begin{flalign}\label{lemma1_end}
&\liminf_{n\rightarrow\infty}\mathcal{R}_{P,n}=\liminf_{n\rightarrow\infty}\mathcal{R}_{Q,n}\nonumber\\
&\limsup_{n\rightarrow\infty}\mathcal{R}_{P,n}=\limsup_{n\rightarrow\infty}\mathcal{R}_{Q,n}.
\end{flalign}
\end{lemma}

\begin{IEEEproof}
\begin{flalign}\label{lemma_discard_eq_first}
&I_{P_{\theta}}(\textbf{X},\textbf{Y};\textbf{Z})\nonumber\\
&=I_{P_{\theta}}(\textbf{X},\textbf{Y};Z_{D_1+1}^{n-D_2})+I_{P_{\theta}}(\textbf{X},\textbf{Y};Z^{D_1}|Z_{D_1+1}^{n-D_2})\nonumber\\
&+I_{P_{\theta}}(\textbf{X},\textbf{Y};Z_{n-D_2+1}^{n}|Z^{n-D_2})\nonumber\\
&\leq I_{P_{\theta}}(\textbf{X},\textbf{Y};Z_{D_1+1}^{n-D_2})+H_{P_{\theta}}(Z^{D_1})+H_{P_{\theta}}(Z_{n-D_2+1}^{n})\nonumber\\
&\leq I_{P_{\theta}}(\textbf{X},\textbf{Y};Z_{D_1+1}^{n-D_2})+(D_1+D_2)\log|\mathcal{Z}|\nonumber\\
&\stackrel{(a)}{\leq} I_{Q_{\theta}}(\textbf{X},\textbf{Y};\textbf{Z})+(D_1+D_2)\log|\mathcal{Z}|.
\end{flalign}
where (a) follows since
\begin{flalign}
&I_{P_{\theta}}(\textbf{X},\textbf{Y};Z_{D_1+1}^{n-D_2})=I_{Q_{\theta}}(\textbf{X},\textbf{Y};Z_{D_1+1}^{n-D_2}).
\end{flalign}
Similarly,
\begin{flalign}
&I_{P_{\theta}}(\textbf{Y};\textbf{Z}|\textbf{X})\leq I_{P_{\theta}}(\textbf{Y};Z_{D_1+1}^{n-D_2}|\textbf{X})+(D_1+D_2)\log|\mathcal{Z}|\nonumber\\
&\stackrel{(a)}{\leq}  I_{Q_{\theta}}(\textbf{Y};\textbf{Z}|\textbf{X})+(D_1+D_2)\log|\mathcal{Z}|,\nonumber\\
&I_{Q_{\theta}}(\textbf{X},\textbf{Y};\textbf{Z})\leq I_{Q_{\theta}}(\textbf{X},\textbf{Y};Z_{D_1+1}^{n-D_2})+(D_1+D_2)\log|\mathcal{Z}|\nonumber\\
&\stackrel{(b)}{\leq}  I_{P_{\theta}}(\textbf{X},\textbf{Y};\textbf{Z})+(D_1+D_2)\log|\mathcal{Z}|,\nonumber\\
&I_{Q_{\theta}}(\textbf{Y};\textbf{Z}|\textbf{X}) \leq I_{Q_{\theta}}(\textbf{Y};Z_{D_1+1}^{n-D_2}|\textbf{X})+(D_1+D_2)\log|\mathcal{Z}|\nonumber\\
&\stackrel{(c)}{\leq}  I_{P_{\theta}}(\textbf{Y};\textbf{Z}|\textbf{X})+(D_1+D_2)\log|\mathcal{Z}|,
\end{flalign}
where (a) and (c) follows since
\begin{flalign}
&I_{P_{\theta}}(\textbf{Y};Z_{D_1+1}^{n-D_2}|\textbf{X})=I_{Q_{\theta}}(\textbf{Y};Z_{D_1+1}^{n-D_2}|\textbf{X}),
\end{flalign}
and (b) follows since
\begin{flalign}
&I_{P_{\theta}}(\textbf{X},\textbf{Y};Z_{D_1+1}^{n-D_2})=I_{Q_{\theta}}(\textbf{X},\textbf{Y};Z_{D_1+1}^{n-D_2}).
\end{flalign}
Consequently,
\begin{flalign}\label{lemma_discard_eq_last}
&\mathcal{R}_{P,n}\subset\mathcal{R}_{Q,n}+\frac{1}{n}(D_1+D_2)\log|\mathcal{Z}|\cdot U\nonumber\\
&\mathcal{R}_{Q,n}\subset\mathcal{R}_{P,n}+\frac{1}{n}(D_1+D_2)\log|\mathcal{Z}|\cdot U
\end{flalign}
where $U$ is the unit square $\{(x,y):0\leq x \leq1,0\leq y \leq1\}$.\

\noindent One can see that Equation (\ref{lemma_discard_eq_last}) results in,
\begin{flalign}
&\liminf_{n\rightarrow\infty}\mathcal{R}_{P,n}=\liminf_{n\rightarrow\infty}\mathcal{R}_{Q,n},\nonumber\\
&\limsup_{n\rightarrow\infty}\mathcal{R}_{P,n}=\limsup_{n\rightarrow\infty}\mathcal{R}_{Q,n}.
\end{flalign}
\end{IEEEproof}

\subsection{Proof of Theorem~\ref{ACMAC_capacity}}\label{ACMAC_C1}

\begin{IEEEproof}[Achievability]
Let $n$ be the transmission block length, i.e., the length of the codewords $\textbf{x}_1,\textbf{x}_2$, and let $\tilde{n}\triangleq \left\lfloor \frac{n}{k}\right\rfloor$.
Denote by $\tilde{x}_i$ the hyper-symbol of length $k\geq D$ that consists of the $k$ consecutive symbols
$(x_{(i-1)\cdot k+1},\ldots,x_{i\cdot k})$ of a vector $\textbf{x}$ of length $n$.

Partitioning each vector $\textbf{x}_1,\textbf{x}_2$  into vectors that consist of hyper-symbols of length $k$, yields the vectors
 \begin{flalign}
 \tilde{\textbf{x}}_1&=(\tilde{x}_{1,1},\ldots,\tilde{x}_{1,\tilde{n}}),\nonumber\\
 \tilde{\textbf{x}}_2&=(\tilde{x}_{2,1},\ldots,\tilde{x}_{2,\tilde{n}}),
 \end{flalign}
 where $\tilde{x}_{1,i}\in\mathcal{X}_1^k$,
 $\tilde{x}_{2,i}\in\mathcal{X}_2^k$ for all $i\in \{1,\ldots,\tilde{n}\}$.

\textbf{Codebook Generation:}
The codebooks $\mathcal{C}^{(1)}$ and $\mathcal{C}^{(2)}(l), 1 \leq l\leq |\mathcal{C}^{(1)}|$ are produced in the following manner:

Set $P_{\tilde{X}_1},P_{\tilde{X}_2|\tilde{X}_1}$, and fix the rates $\tilde{R}_1$ and $\tilde{R}_2$. Let $\mathcal{C}^{(1)}$ be the codebook of the common message $M_1$, which consists of $2^{\tilde{n}\tilde{R}_1}$ codewords, each of these codewords  is generated  according to $P(\tilde{\textbf{x}}_1)=\prod_{i=1}^{\tilde{n}} P_{\tilde{X}_1}(\tilde{x}_{1,i})$.

For every $\tilde{\textbf{x}}_1(l)\in\mathcal{C}^{(1)}$ generate randomly and independently $2^{\tilde{n}\tilde{R}_2}$ codewords  $\{\tilde{\textbf{x}}_2(l,1),\ldots,\tilde{\textbf{x}}_2(l,2^{\tilde{n}\tilde{R}_2})\}$ according to
$P(\tilde{\textbf{x}}_2|\tilde{\textbf{x}}_1)=\prod_{i=1}^{\tilde{n}} P_{\tilde{X}_2|\tilde{X}_1}(\tilde{x}_{2,i}|\tilde{x}_{1,i})$. We denote
$\{\tilde{\textbf{x}}_2(l,1),\ldots,\tilde{\textbf{x}}_2(l,2^{\tilde{n}\tilde{R}_2})\}$ by $\mathcal{C}^{(2)}(l)$.

\textbf{Encoding:}
To send the messages $m_1,m_2$, encoder 1 sends $\tilde{\textbf{x}}_1(m_1)$, and encoder 2 sends $\tilde{\textbf{x}}_2(m_1,m_2)$.

\textbf{Decoding:}
Denote by $\overline{y}_i$ the hyper-symbol $(y_{(i-1)\cdot k+1+d_{max}},\ldots,y_{i\cdot k-d_{min}})$ that is, $\overline{y}_i\in\mathcal{Y}^{k-D+1}$. Further, let $\overline{\textbf{y}}$ denote the vector  $(\overline{y}_{1},\ldots,\overline{y}_{\tilde{n}})$.
Suppose that the actual delay in the channel is $d\in\mathcal{D}$. Denote the set of all vectors
$(\tilde{\textbf{x}}_1$,$\tilde{\textbf{x}}_2,\overline{\textbf{y}})\in \mathcal{X}_1^{\tilde{n}\cdot k}\times\mathcal{X}_2^{\tilde{n}\cdot k}\times \mathcal{Y}^{\tilde{n}\cdot(k-D+1)}$
that are $\epsilon$-strongly typical with respect to a p.m.f. $P_d(\tilde{x}_1,\tilde{x}_2,\overline{y})$ by $T_{d,\epsilon}^n(\tilde{X}_1,\tilde{X}_2,\overline{Y})$,
where
\begin{flalign}
&P_d(\tilde{x}_1,\tilde{x}_2,\overline{y})=P_{\tilde{X}_1}(\tilde{x}_1)P_{\tilde{X}_2|\tilde{X}_1}(\tilde{x}_2|\tilde{x}_1)P_d(\overline{y}|\tilde{x}_1,\tilde{x}_2),\nonumber\\
&P_d(\overline{y}|\tilde{x}_1,\tilde{x}_2)=
\prod_{j=d_{max}+1}^{k-d_{min}} P_{Y|X_1,X_2}(\overline{y}_{j-d_{max}}|\tilde{x}_{1,j-d},\tilde{x}_{2,j}),\nonumber\\
&P_d(\tilde{\textbf{x}}_1,\tilde{\textbf{x}}_2,\overline{\textbf{y}})=\prod_{i=1}^{\tilde{n}}P_d(\tilde{x}_{1,i},\tilde{x}_{2,i},\overline{y}_i).
\end{flalign}
We remark that we defined $\overline{y}_i$ in such a manner that given $\tilde{\textbf{x}}_1,\tilde{\textbf{x}}_2$, $\overline{y}_i$ is independent of  $\overline{y}_1^{i-1}$ and $\overline{y}_{i+1}^n$.

Given that the decoder knows the delay $d$, it looks for $\hat{m}_1\in\{1,\ldots,2^{nR_1}\}$ and $\hat{m}_2\in\{1,\ldots,2^{nR_2}\}$ such that
\begin{flalign}
(\tilde{\textbf{x}}_1(\hat{m}_1),\tilde{\textbf{x}}_2(\hat{m}_2),\overline{\textbf{y}})\in T_{d,\epsilon}^n(\tilde{X}_1,\tilde{X}_2,\overline{Y}).
\end{flalign}
\textbf{Analysis of Probability of Error:}
Suppose that the pair of messages $(m_1,m_2)=(1,1)$ is sent, and that the delay is $d\in\mathcal{D}$. An error is made if one or more of the following events occur:
\begin{flalign}
&\mathcal{E}_1=\{\tilde{\textbf{x}}_1(1)\notin T_{\epsilon}^{\tilde{n}}(\tilde{X}_1)\}\nonumber\\
&\mathcal{E}_2=\{(\tilde{\textbf{x}}_2(1,1),\tilde{\textbf{x}}_1(1))\notin T_{\epsilon}^{\tilde{n}}(\tilde{X}_2,\tilde{X}_1)\}\nonumber\\
&\mathcal{E}_3=
\begin{Bmatrix}
(\tilde{\textbf{x}}_1(1),\tilde{\textbf{x}}_2(1,1),\overline{\textbf{y}})\notin T_{d,\epsilon}^{\tilde{n}}(\tilde{X}_1,\tilde{X}_2,\overline{Y})
\end{Bmatrix}\nonumber\\
&\mathcal{E}_4=\begin{Bmatrix}
\exists \hat{m}_1\neq1
\text{ and } \hat{m}_2\in\{1,\ldots,2^{\tilde{n}\tilde{R}_2}\}  \text{ s.t. }\\ (\tilde{\textbf{x}}_1(\hat{m}_1),\tilde{\textbf{x}}_2(\hat{m}_1,\hat{m}_2),\overline{\textbf{y}})\in T_{d,\epsilon}^{\tilde{n}}(\tilde{X}_1,\tilde{X}_2,\overline{Y})
\end{Bmatrix}\nonumber\\
&\mathcal{E}_5=\begin{Bmatrix}
\exists \hat{m}_2\neq1 \text{ s.t. } \\
(\tilde{\textbf{x}}_1(1),\tilde{\textbf{x}}_2(1,\hat{m}_2),\overline{\textbf{y}})\in T_{d,\epsilon}^{\tilde{n}}(\tilde{X}_1,\tilde{X}_2,\overline{Y})
\end{Bmatrix}.
\end{flalign}
By the union bound,
\begin{flalign}
\Pr(\mathcal{E})&=\Pr\left(\bigcup_{i=1}^5 \mathcal{E}_i\right)\leq \Pr(\mathcal{E}_1)+\Pr(\mathcal{E}_1^c\cap \mathcal{E}_2)+\Pr(\mathcal{E}_2^c\cap \mathcal{E}_3)
+\Pr(\mathcal{E}_4)+\Pr(\mathcal{E}_5).
\end{flalign}

First, by the law of large numbers (LLN)  $\Pr(\mathcal{E}_1)\rightarrow 0$ as $n\rightarrow\infty$.
Second, the conditional typicality lemma \cite[p. 27]{NetworkInformationTheory} dictates that
$\Pr(\mathcal{E}_1^c\cap \mathcal{E}_2)\rightarrow 0$ as $\tilde{n}\rightarrow\infty$.

The sequence $\overline{\textbf{y}}$ is generated given $\tilde{\textbf{x}}_1$ and $\tilde{\textbf{x}}_2$ according to $\prod_{i=1}^{\tilde{n}}P_d(\overline{y}_{i}|\tilde{x}_{1,i},\tilde{x}_{2,i})$, therefore, from the LLN we have that $\Pr(\mathcal{E}_2^c\cap \mathcal{E}_3)$ vanishes as $\tilde{n}$ tends to infinity.

Finally, by the packing lemma \cite[p. 46]{NetworkInformationTheory}, $\Pr(\mathcal{E}_4)\rightarrow0$ as $\tilde{n}\rightarrow\infty$ if
\begin{flalign}\label{eq:cap_up_1}
k(R_1+R_2)=\tilde{R}_1+\tilde{R}_2 \leq I_d(\tilde{X}_1,\tilde{X}_2;\overline{Y}),
\end{flalign}
and $\Pr(\mathcal{E}_5)\rightarrow0$ as $\tilde{n}\rightarrow\infty$ if
\begin{flalign}\label{eq:cap_up_2}
kR_2=\tilde{R}_2 \leq I_d(\tilde{X}_2;\overline{Y}|\tilde{X}_1).
\end{flalign}

Let $\textbf{x}_1\in\mathcal{X}_1^k $ and  $\textbf{x}_2 \in\mathcal{X}_2^k$. Denote for every $k$, $P(\textbf{x}_1,\textbf{x}_2)$, and $d\in\mathcal{D}$,
\begin{flalign}
&\mathcal{Q}_{k,d}\left(P(\textbf{x}_1,\textbf{x}_2)\right)=
 \left\{
\begin{array}{ll}
    \multirow{2}{*}{$(R_1,R_2):$}& R_1 +R_2  \leq \frac{1}{k}I_d(\textbf{V}_1,\textbf{X}_2;Y_{d_{max}+1}^{k-d_{min}}),\\
    &\hspace{1cm}R_2  \leq \frac{1}{k}I_d(\textbf{X}_2;Y_{d_{max}+1}^{k-d_{min}}|\textbf{X}_1)
\end{array}
\right\}.
\end{flalign}

Since the encoder does not know the delay  $d\in\mathcal{D}$,  a rate-pair is achievable given $P(\textbf{x}_1,\textbf{x}_2)$ if it lies in the intersection of all the regions $\underline{R}_{k,d}(P(\textbf{x}_1,\textbf{x}_2))$. Therefore, by (\ref{eq:cap_up_1}) and (\ref{eq:cap_up_2})
\begin{flalign}
\mathcal{Q}_k(P(\textbf{x}_1,\textbf{x}_2))=\bigcap_{d\in\mathcal{D}}
\mathcal{Q}_{k,d}(P(\textbf{x}_1,\textbf{x}_2))
\end{flalign}
is an achievable rate region.

Consequently, the closure of the region $\bigcup_{k \geq D} \mathcal{Q}_k$ is achievable.

\end{IEEEproof}

\begin{IEEEproof}[Converse]
Let,
\begin{flalign}
P_d(m_1,m_2,\textbf{x}_1,\textbf{x}_2,\textbf{y})=P(m_1)P(m_2)P(\textbf{x}_1|m_1)P(\textbf{x}_2|\textbf{x}_1)P_d(\textbf{y}|\textbf{x}_1,\textbf{x}_2),
\end{flalign}
where
\begin{flalign}
&P(m_1)=2^{-nR_1},\quad P(m_2)=2^{-nR_2}\newline\\
&P_d(\textbf{y}|\textbf{x}_1,\textbf{x}_2)=\prod_{i=1}^nP(y_i|x_{1,i-d},x_{2,i}).
\end{flalign}
We denote Information-Theoretic functionals of $P_d(m_1,m_2,\textbf{x}_1,\textbf{x}_2,\textbf{y})$ by the subscript $d$, e.g., $H_d(M_1,M_2|\textbf{Y})$.

We first upper bound the sum-rate $R_1+R_2$. Let $\delta_n=\frac{1}{n}H_d(M_1,M_2|\textbf{Y})$.
For every sequence of $(2^{nR_1},2^{nR_2},n)$-codes with probability of error $P_e^{(n)}$ that vanishes as $n$ tends to infinity for every $d\in \mathcal{D}$,
\begin{flalign}
n(R_1+R_2) &= H(M_1,M_2)\nonumber\\
&= H(M_1,M_2) - H_d(M_1,M_2|\textbf{Y})+H_d(M_1,M_2|\textbf{Y}) \nonumber\\
&=  I_d(M_1,M_2;\textbf{Y})+n\delta_n \nonumber\\
&=  H_d(\textbf{Y})-H_d(\textbf{Y}|M_1,M_2)+n\delta_n \nonumber\\
&\stackrel{(a)}{=}  H_d(\textbf{Y})-H_d(\textbf{Y}|M_1,M_2,\textbf{X}_1,\textbf{X}_2)+n\delta_n \nonumber\\
&\stackrel{(b)}{=}  H_d(\textbf{Y})-H_d(\textbf{Y}|\textbf{X}_1,\textbf{X}_2)+n\delta_n \nonumber\\
&=  I_d(\textbf{X}_1,\textbf{X}_2;\textbf{Y})+n\delta_n
\end{flalign}
where (a) follows since $\textbf{X}_1$ is a function of $M_1$, $\textbf{X}_2$ is a function of $M_1$ and $\textbf{X}_1$, and from the fact that conditioning reduces entropy, and (b) follows from the fact that $(M_1,M_2)-(\textbf{X}_1,\textbf{X}_2,d)-\textbf{Y}$ is a Markov chain. Additionally, since the error probability $P_e^{(n)}$  vanishes as $n$ tends to infinity, Fano's Inequality yields that $\delta_n$ vanishes as $n$ tends to infinity.
\begin{flalign}
nR_2 &= H(M_2|M_1)\nonumber\\
&= H(M_2|M_1) - H_d(M_2|M_1,\textbf{Y})
+H_d(M_2|M_1,\textbf{Y}) \nonumber\\
&\stackrel{(a)}{\leq} I_d(M_2;\textbf{Y}|M_1)+n\delta_n\nonumber\\
&=  H_d(\textbf{Y}|M_1)-H_d(\textbf{Y}|M_1,M_2)+n\delta_n\nonumber\\
&\stackrel{(b)}{=}  H_d(\textbf{Y}|M_1,\textbf{X}_1)
-H_d(\textbf{Y}|M_1,M_2,\textbf{X}_1,\textbf{X}_2)+n\delta_n\nonumber\\
&\stackrel{(c)}{=}  H_d(\textbf{Y}|\textbf{X}_1)-H_d(\textbf{Y}|\textbf{X}_1,\textbf{X}_2)+n\delta_n\nonumber\\
&=  I_d(\textbf{X}_2;\textbf{Y}|\textbf{X}_1)+n\delta_n
\end{flalign}
where $(a)$ follows from the definition of $\delta_n$ and by nonnegativity of the entropy,  (b) follows since $\textbf{X}_1$ is a function of $M_1$, $\textbf{X}_2$ is a function of $M_2$ and $\textbf{X}_1$, and from the fact that conditioning reduces entropy, and (c) follows from the fact that $(M_1,M_2)-(\textbf{X}_1,\textbf{X}_2,d)-\textbf{Y}$ and $M_1-(\textbf{X}_1,d)-\textbf{Y}$ are a Markov chains.

Since both encoders do not know the delay $d$, $\textbf{X}_1$ and $\textbf{X}_2$ do not depend on the delay $d$.
Therefore, we can write the following outer rate region as a function of $P(\textbf{x}_1,\textbf{x}_2)$, $d\in\mathcal{D}$ and $n$,
\begin{flalign}\label{MAC_upper_0}
&\mathcal{R}_{n,d}\left(P(\textbf{x}_1,\textbf{x}_2)\right)=
\left\{
\begin{array}{ll}
    \multirow{2}{*}{$(R_1,R_2):$}&R_1+R_2 \leq \frac{1}{n}I_d(\textbf{X}_1,\textbf{X}_2;\textbf{Y})+\delta_n\\
&\hspace{1cm}R_2 \leq \frac{1}{n}I_d(\textbf{X}_2;\textbf{Y}|\textbf{X}_1)+\delta_n
\end{array}
\right\}.
\end{flalign}

In addition, the fact that both encoders do not know the delay $d$  means that a rate-pair is achievable only if it lies in the intersection over $d$, of all the regions $R_{n,d}\left(P(\textbf{x}_1,\textbf{x}_2)\right)$. Denote
\begin{flalign}
\mathcal{R}_n\left(P(\textbf{x}_1,\textbf{x}_2)\right)=\bigcap_{d\in\mathcal{D}}
\mathcal{R}_{n,d}\left(P(\textbf{x}_1,\textbf{x}_2)\right).
\end{flalign}
The union over all p.m.f. $P(\textbf{x}_1,\textbf{x}_2)$ yields the outer rate region
\begin{flalign}
\mathcal{R}_n=\bigcup_{P(\textbf{x}_1,\textbf{x}_2)}\mathcal{R}_n\left(P(\textbf{x}_1,\textbf{x}_2)\right).
\end{flalign}
Finally, taking $n\rightarrow\infty$, and noting that $\delta_n$ vanishes as $n$ tends to infinity, yields that the capacity region is included in the region
\begin{flalign}
&\liminf_{n\rightarrow\infty}\bigcup_{P(\textbf{x}_1,\textbf{x}_2)}\bigcap_{d\in\mathcal{D}}
\left\{
\begin{array}{ll}
    \multirow{2}{*}{$(R_1,R_2):$}&R_1+R_2 \leq \frac{1}{n}I_d(\textbf{X}_1,\textbf{X}_2;\textbf{Y})\\
&\hspace{1cm}R_2 \leq \frac{1}{n}I_d(\textbf{X}_2;\textbf{Y}|\textbf{X}_1)
\end{array}
\right\}.
\end{flalign}
\end{IEEEproof}

Now, we finish the proof of Theorem \ref{ACMAC_capacity}. By the achievability part, the region $\bigcup_{n \geq D} \mathcal{Q}_n$ is achievable.
By definition
\begin{flalign}
\lim\inf_{n\rightarrow\infty}\mathcal{Q}_n \subseteq \lim\sup_{n\rightarrow\infty}\mathcal{Q}_n
\subseteq \bigcup_{n \geq D} \mathcal{Q}_n,
\end{flalign}
by Lemma~\ref{lemma_discard} (see (\ref{lemma1_begin})-(\ref{lemma1_end})), it follows that:
\begin{flalign}
&\lim\inf_{n\rightarrow\infty}\mathcal{Q}_n=\lim\inf_{n\rightarrow\infty}\mathcal{R}_n\\
&\lim\sup_{n\rightarrow\infty}\mathcal{Q}_n=\lim\sup_{n\rightarrow\infty}\mathcal{R}_n,
\end{flalign}
where $\mathcal{R}_n$ is defined in (\ref{R_n_def_ACMAC}).
Therefore,
\begin{flalign}
\lim\inf_{n\rightarrow\infty}\mathcal{R}_n \subseteq \lim\sup_{n\rightarrow\infty}\mathcal{R}_n
\subseteq \bigcup_{n \geq D} \mathcal{Q}_n
\end{flalign}
are achievable rate regions.
Furthermore, by the converse part the region $\liminf_{n\rightarrow\infty}\mathcal{R}_n$ is an outer bound on the capacity region of the AC-MAC. Therefore,
\begin{flalign}
\mathcal{C}=\text{closure}\left(\bigcup_{n \geq D} \mathcal{Q}_n\right)=
\text{closure}(\lim\sup_{n\rightarrow\infty}\mathcal{R}_n)
=\text{closure}(\lim\inf_{n\rightarrow\infty}\mathcal{R}_n).
\end{flalign}

\subsection{Proof of Theorem~\ref{ACMAC_inner_single}}\label{ACMAC_A2}

As mentioned before, by sending predefined training sequences in the first $o(n)$ bits, the decoder can deduce the delay with probability of error that vanishes as $n$ tends to infinity. Therefore, we can assume that the decoder knows the delay $d$.
In addition, we ignore the end effects in our notations, since the first/last symbols do not affect the asymptotic performance in terms of the reliably transmitted rates.

\textbf{Codebook Generation:}
The codebooks $\mathcal{C}^{(1)},\mathcal{C}^{(1')}$ and $\mathcal{C}^{(2)}(l), 1 \leq l\leq |\mathcal{C}^{(1)}|$ are produced in the following manner:

Set $P_{X_1}(x_1),P_{X_2|\overline{V}}(x_2|\overline{v})$. Let $\mathcal{C}^{(1)}$ be the codebook of the common message $M_1$, which consists of $2^{nR_1}$ codewords, each of these codewords  is generated  according to $P(\textbf{x}_1)=\prod_{i=1}^n P_{X_1}(x_{1,i})$.

The codebook $\mathcal{C}^{(1')}$ is produced from $\mathcal{C}^{(1)}$ by following the one-to-one mapping: let $\textbf{x}_1(l)$ be the $l$th
codeword in  $\mathcal{C}^{(1)}$, then for every $i\in \{1,\ldots,n\}$ define
\begin{flalign}\label{x_to_v}
\overline{v}_i(l)=(x_{1,i-d_{max}}(l),\ldots,x_{1,i+d_{min}}(l)).
\end{flalign}
The resulting codeword $\textbf{v}(l)=(\overline{v}_1(l),\ldots,\overline{v}_n(l))$ is the $l$-th codeword in $C^{(1')}$,
that is, the codewords in $\mathcal{C}^{(1)}$ appear in $\mathcal{C}^{(1')}$ as vectors that were produced by a sliding window of size $D$ on the sequence $\textbf{x}_1$.

Now, for every $\textbf{v}(l)\in\mathcal{C}^{(1')}$ generate randomly and independently $2^{nR_2}$
codewords\newline   $\{\textbf{x}_2(l,1),\ldots,\textbf{x}_2(l,2^{nR_2})\}$ according to
$P(\textbf{x}_2|\textbf{v})=\prod_{i=1}^n P_{X_2|\overline{V}}(x_{2,i}|\overline{v}_i)$.

We denote
$\{\textbf{x}_2(l,1),\ldots,\textbf{x}_2(l,2^{nR_2})\}$ by $\mathcal{C}^{(2)}(l)$.

\textbf{Encoding:}
To send the messages $m_1,m_2$ encoder 1 sends $\textbf{x}_1(m_1)$ and encoder 2 sends $\textbf{x}_2(m_1,m_2)$.

\textbf{Decoding:}
We define the following function to align the uninformed encoder's codeword with the output given the delay in the channel.
Suppose that the actual delay in the channel is $d\in\mathcal{D}$. Let $\sigma(\textbf{x}_1,d)$ be the function
\begin{flalign} \label{sigma_def}
\sigma(\textbf{x}_1,d)=\begin{cases}
(x_{1,n-d+1},\ldots,x_{1,n},x_{1,1},\ldots,x_{1,n-d})\hspace{-0.2cm}&\text{if } d\geq 0 \\
(x_{1,|d|},\ldots,x_{1,n},x_{1,1},\ldots,x_{1,|d|-1})\hspace{-0.2cm}&\text{if } d< 0.
\end{cases}
\end{flalign}
Given that the decoder knows the delay $d$, it first looks for $\hat{m}_1\in\{1,\ldots,2^{nR_1}\}$ such that
\begin{flalign}
(\sigma(\textbf{x}_1(\hat{m}_1),d),\textbf{y})\in T_{d,\epsilon}^n(X_1,Y)
\end{flalign}
where
\begin{flalign}
P_d(x_1,y)=\sum_{\overline{v},x_2}P(\overline{v})\mathbbm{1}_{\{v_{d_{max}-d+1}=x_1\}}P(x_2|\overline{v})P(y|x_2,x_1),
\end{flalign}
and $T_{d,\epsilon}^n(X_1,Y)$ is the set of all vectors $(\textbf{x}_1,\textbf{y})$ that are $\epsilon$-strongly typical with respect to $P_d(x_1,y)$.

Once the decoder recovers the sequence $\textbf{x}_1(\hat{m}_1)$, it can deduce the sequence $\textbf{v}(\hat{m}_1)$ by the one-to-one mapping which is stated by Eq. (\ref{x_to_v}).  Then, with the delay knowledge that, as mentioned before, exists at the decoder, it looks for $\hat{m}_2\in\{1,\ldots,2^{nR_2}\}$ such that
\begin{flalign}
(\textbf{v}(\hat{m}_1),\textbf{x}_2(\hat{m}_1,\hat{m}_2),\textbf{y})\in T_{d,\epsilon}^n(\overline{V},X_2,Y)
\end{flalign}
where
\begin{flalign}
P_d(\overline{v},x_2,y)=P(\overline{v})P(x_2|\overline{v})P(y|x_2,v_{d_{max}-d+1}),
\end{flalign}
and $T_{d,\epsilon}^n(\overline{V},X_2,Y)$ is the set of all vectors $(\textbf{v},\textbf{x}_2,\textbf{y})$ that are $\epsilon$-strongly typical with respect to $P_d(\overline{v},x_2,y)$.

\textbf{Analysis of the probability of error:}
Suppose that the pair of messages $(m_1,m_2)=(1,1)$ is sent, and that the delay is $d\in\mathcal{D}$. An error is made if one or more of the following events occur:
\begin{flalign}
&\mathcal{E}_1=\{\textbf{x}_1(1)\notin T_{\epsilon}^n(X_1)\}\nonumber\\
&\mathcal{E}_2=\{\textbf{v}(1)\notin T_{\epsilon}^n(\overline{V})\}\nonumber\\
&\mathcal{E}_3=\{(\textbf{x}_2(1,1),\textbf{v}(1))\notin T_{\epsilon}^n(X_2,\overline{V})\}\nonumber\\
&\mathcal{E}_4=\begin{Bmatrix}(\sigma(\textbf{x}_1(1),d),\textbf{y})\notin T_{d,\epsilon}^n(X_1,Y) \text{ or }\\(\textbf{v}(1),\textbf{x}_2(1,1),\textbf{y})\notin T_{d,\epsilon}^n(\overline{V},X_2,Y)\end{Bmatrix}\nonumber\\
&\mathcal{E}_5=\{\exists \hat{m}_1\neq1 \text{ s.t. }(\sigma(\textbf{x}_1(\hat{m}_1),d),\textbf{y})\in T_{d,\epsilon}^n(X_1,Y)\}\nonumber\\
&\mathcal{E}_6=\{\exists \hat{m}_2\neq1 \text{ s.t. }(\textbf{x}_2(1,\hat{m}_2),\textbf{y})\in T_{d,\epsilon}^n(X_2,Y|\textbf{v}(1))\}
\end{flalign}
where the function $\sigma(\cdot,\cdot)$ is defined in (\ref{sigma_def}).

We remark that due to the notation of typical sets, we define $\sigma(\cdot,\cdot)$  in such a manner that given the delay $d$, the vector $\sigma(\textbf{x}_1,d)$ is aligned with the output vector $\textbf{y}$. We note that we do not need such a notation for $\epsilon_6$ since at each time instant $i$ the delayed input $x_{1,i-d}$ is  part of the vector $\overline{v}_i$.

By the union bound,
\begin{flalign}
\Pr(\mathcal{E})&=\Pr\left(\bigcup_{i=1}^6 \mathcal{E}_i\right)\leq \Pr(\mathcal{E}_1)+\Pr(\mathcal{E}_1^c\cap \mathcal{E}_2)+\Pr(\mathcal{E}_2^c\cap \mathcal{E}_3) + \Pr(\mathcal{E}_3^c\cap \mathcal{E}_4)+\Pr(\mathcal{E}_5)+\Pr(\mathcal{E}_6).
\end{flalign}

From the LLN $\Pr(\mathcal{E}_1)\rightarrow0$ as $n\rightarrow\infty$.
In addition, from the stationarity and ergodicity of $v^{n}$ we infer that $\Pr(\mathcal{E}_2)\rightarrow 0$ as $n\rightarrow\infty$.

By the  conditional typicality lemma \cite[p. 46]{NetworkInformationTheory}  $\Pr(\mathcal{E}_3)\rightarrow 0$ as $n\rightarrow\infty$. Additionally, the conditional typicality lemma \cite[p. 46]{NetworkInformationTheory}  implies that $\Pr(\mathcal{E}_4)\rightarrow 0$ as $n\rightarrow\infty$.

Since all $\textbf{x}_1\in \mathcal{C}^{(1)}$ were generated according to an i.i.d. distribution, we can use the packing lemma \cite[p. 46]{NetworkInformationTheory} to deduce that  $\Pr(\mathcal{E}_5)\rightarrow 0$ as $n\rightarrow\infty$ if
\begin{flalign}
R_1 < I_d(X_1;Y).
\end{flalign}

Now we notice that,
\begin{flalign}
P_d(\textbf{y}|\textbf{x}_2,\tilde{\textbf{v}})=\prod_{i=1}^n P(y_i|x_{2,i},\overline{v}_{i,d_{max}-d+1})
\end{flalign}
that is, $\textbf{y}$ is memoryless given the sequences $\textbf{x}_2,\tilde{\textbf{v}}$ and the delay. Therefore, an additional use of the packing lemma yields
$\Pr(\mathcal{E}_6)\rightarrow 0$ as $n\rightarrow\infty$ if
\begin{flalign}
R_2 < I_d(X_2;Y|\overline{V}).
\end{flalign}

We can argue that if $(R_1,R_2)$ is an achievable rate, then $(R_1+R_2,0)$ is an achievable rate as well. This is true since we can decompose each common message $m_1$ into two common sub-messages $(m_{1_1},m_{1_2})$ and let encoder $1$ send $m_{1_1}$ and encoder $2$ send $m_{1_2}$ in addition to $m_2$. The decoder finds $(m_{1_1},m_{1_2})$ and can assemble the message $m_1$.

Therefore, we can write the following rate region for every $P(x_1),P(x_2|\overline{v})$, and $d\in\mathcal{D}$,
\begin{flalign}
&\underline{\mathcal{R}}_d(P(x_1),P(x_2|\overline{v}))=
 \left\{
\begin{array}{ll}
    \multirow{2}{*}{$(R_1,R_2):$}& R_1 +R_2 \leq I_d(X_1;Y)+I_d(X_2;Y|\overline{V}),\\
    &\hspace{1cm}R_2 \leq I_d(X_2;Y|\overline{V})
\end{array}
\right\}.
\end{flalign}

Since the encoder does not know the delay  $d\in\mathcal{D}$,  a rate-pair is achievable for fixed $P(x_1),P(x_2|\overline{v})$ if it lies in the intersection of all the regions $\underline{R}_d(P(x_1),P(x_2|\overline{v}))$. Therefore,
\begin{flalign}
\underline{\mathcal{R}}(P(x_1),P(x_2|\overline{v}))&=\bigcap_{d\in\mathcal{D}}
\underline{\mathcal{R}}_d(P(x_1),P(x_2|\overline{v}))\nonumber\\
& =\left\{
\begin{array}{ll}
    \multirow{2}{*}{$(R_1,R_2):$}& R_1 +R_2 \leq \min_{d\in\mathcal{D}}\left[I_d(X_1;Y)+I_d(X_2;Y|\overline{V})\right],\\
    &\hspace{1cm}R_2 \leq \min_{d\in\mathcal{D}}I_d(X_2;Y|\overline{V})
\end{array}
\right\}
.
\end{flalign}
where the last equality follows since the set of all possible delays is finite.

Consequently, the following rate region
\begin{flalign}
\underline{\mathcal{R}}&=\bigcup_{P(x_1),P(x_2|\overline{v})}\underline{\mathcal{R}}(P(x_1),P(x_2|\overline{v})),
\end{flalign}
is achievable.

Finally, since $d_{max},d_{min}<\infty$ we can use time sharing arguments to infer that the closure convex of the rate region $\underline{\mathcal{R}}$ is an achievable rate region for the ACMAC.

\subsection{Proof of Theorem~\ref{ACMAC_outer_single}}\label{ACMAC_C2}

Let,
\begin{flalign}
P_d(m_1,m_2,\textbf{x}_1,\textbf{v},\textbf{x}_2,\textbf{y})=P(m_1)P(m_2)P(\textbf{x}_1|m_1)P(\textbf{v}|\textbf{x}_1)P(\textbf{x}_2|\textbf{x}_1)P_d(\textbf{y}|\textbf{x}_1,\textbf{x}_2),
\end{flalign}
where
\begin{flalign}
&P(m_1)=2^{-nR_1},\quad P(m_2)=2^{-nR_2},\nonumber\\
&P(\textbf{v}|\textbf{x}_1)=\prod_{i=1}^n\mathbbm{1}_{\{v_i=(x_{1,i-d_{max}},\ldots,x_{1,i+d_{min}})\}}\nonumber\\
&P_d(\textbf{y}|\textbf{x}_1,\textbf{x}_2)=\prod_{i=1}^nP(y_i|x_{1,i-d},x_{2,i}).
\end{flalign}
We denote information theoretic functionals of $P_d(m_1,m_2,\textbf{x}_1,\textbf{x}_2,\textbf{y})$ by the subscript $d$, e.g., $H_d(M_1,M_2|\textbf{Y})$.

We first upper bound the sum-rate $R_1+R_2$.
 For every sequence of $(2^{nR_1},2^{nR_2},n)$-codes with probability of error $P_e^{(n)}$ that vanishes as $n$ tends to infinity for every $d\in \mathcal{D}$. Thus, from Fano's inequality  $\delta_n\triangleq n^{-1}H_d(M_1,M_2|\textbf{Y})$ vanishes as well as $n$ tends to infinity. We obtain,\begin{flalign}
n(R_1+R_2) &= H(M_1,M_2)  \nonumber\\
&= H(M_1,M_2) - H_d(M_1,M_2|\textbf{Y})+H_d(M_1,M_2|\textbf{Y}) \nonumber\\
&=  I_d(M_1,M_2;\textbf{Y})+n\delta_n
\end{flalign}
where the second equality follows since the messages of the users do not depend on the delay $d$.

We now bound the term $I_d(M_1,M_2;\textbf{Y})$. Denote $\overline{X}_{1,i}=X_{1,i D+1-d}^{(i+1) D-d}$, $\overline{X}_{2,i}=X_{2,i D+1}^{(i+1) D}$ and $\overline{Y}_i=Y_{i D+1}^{(i+1) D}$. Then,
\begin{flalign}\label{MAC_upper_1}
 &I_d(M_1,M_2;\textbf{Y})  =  H_d(\textbf{Y})-H_{d}(\textbf{Y}|M_1,M_2) \nonumber\\
& =  H_d(\textbf{Y})-H_{d}(\textbf{Y}|M_1,M_2,\textbf{X}_1,\textbf{X}_2) \nonumber\\
&\stackrel{(a)}{=}  H_d(\textbf{Y})-H_{d}(\textbf{Y}|\textbf{X}_1,\textbf{X}_2)\nonumber\\
&= \sum_{i=0}^{n/D-1}  \left[ H_d(\overline{Y}_i|Y^{i\cdot D})-H_{d}(\overline{Y}_i|Y^{i\cdot D},\textbf{X}_1,\textbf{X}_2)\right]\nonumber\\
&\stackrel{(b)}{\leq} \sum_{i=0}^{n/D-1}  \left[ H_d(\overline{Y}_i)-H_{d}(\overline{Y}_i|Y^{i\cdot D},\textbf{X}_1,\textbf{X}_2)\right]\nonumber\\
&\stackrel{(c)}{=} \sum_{i=0}^{n/D-1}  \left[ H_d(\overline{Y}_i)-H_{d}(\overline{Y}_i|\textbf{X}_1,\textbf{X}_2)\right]\nonumber\\
&\stackrel{(d)}{=} \sum_{i=0}^{n/D-1}  \left[ H_d(\overline{Y}_i)-H_{d}(\overline{Y}_i|\overline{X}_{1,i},\overline{X}_{2,i})\right]\nonumber\\
\end{flalign}
where  (a) follows since $M_1,M_2-(\textbf{X}_1,\textbf{X}_2)-\textbf{Y}$ is a Markov chain for any given $d$,
(b)  follows since conditioning reduces entropy and (c) follows since $(\textbf{X}_1,\textbf{X}_2,Y^{iD})-(\overline{X}_{1,i},\overline{X}_{2,i},d)-\overline{Y}_i$ is a Markov chain for any \textsl{given} $d$ and all $i$.
\newline
Thus, we have that
\begin{flalign}
I_d(M_1,M_2;\textbf{Y})&\leq \sum_{i=0}^{n/D}  \left[ H_d(\overline{Y}_i)-H_{d}(\overline{Y}_i|\overline{X}_{1,i},\overline{X}_{2,i})\right]\nonumber\\
&=\sum_{i=0}^{n/D}I_d(\overline{X}_{1,i},\overline{X}_{2,i},\overline{Y}_i)
\end{flalign}

It is left to bound the rate $R_2$.
Let $\delta_n'\triangleq n^{-1}H_d(M_2|M_1,\textbf{Y})$, again, by Fano's Inequality we have that $\delta_n'$ vanishes as $n$ tends to infinity.
 \begin{flalign}
nR_2 &= H(M_2|M_1)  \nonumber\\
&= H(M_2|M_1) - H_d(M_2|M_1,\textbf{Y})+H_d(M_2|M_1,\textbf{Y}) \nonumber\\
&=  I_d(M_2;\textbf{Y}|M_1)+n\delta_n'.
\end{flalign}

We remind the reader that  $\overline{V}_i=(X_{1,i-d_{max}},\ldots,X_{1,i+d_{min}})$ as  defined previously, also let  $\tilde{V}_{i}=\overline{V}_{i D+1-d}^{(i+1) D-d}$. Then,
 \begin{flalign}\label{MAC_upper_2}
& I_d(M_2;\textbf{Y}|M_1) =  H_d(\textbf{Y}|M_1)-H_{d}(\textbf{Y}|M_1,M_2) \nonumber\\
&\stackrel{(a)}{=}  H_d(\textbf{Y}|M_1,\textbf{X}_1)-H_{d}(\textbf{Y}|M_1,M_2,\textbf{X}_1,\textbf{X}_2) \nonumber\\
&\stackrel{(b)}{=}   H_d(\textbf{Y}|M_1,\textbf{X}_1)-H_{d}(\textbf{Y}|\textbf{X}_1,\textbf{X}_2) \nonumber\\
&\stackrel{(c)}{\leq}  H_d(\textbf{Y}|\textbf{X}_1)-H_{d}(\textbf{Y}|\textbf{X}_1,\textbf{X}_2)\nonumber\\
&= \sum_{i=0}^{n/D-1}  \left[ H_d(\overline{Y}_i|Y^{i\cdot D},\textbf{X}_1)-H_{d}(\overline{Y}_i|Y^{i\cdot D},\textbf{X}_1,\textbf{X}_2)\right]\nonumber\\
&\stackrel{(d)}{\leq}\sum_{i=0}^{n/D-1}  \left[ H_d(\overline{Y}_i|\tilde{V}_{i})-H_{d}(\overline{Y}_i|Y^{i\cdot D},\textbf{X}_1,\textbf{X}_2)\right]\nonumber\\
&\stackrel{(e)}{=}\sum_{i=0}^{n/D-1}  \left[ H_d(\overline{Y}_i|\tilde{V}_{i})-H_{d}(\overline{Y}_i|\tilde{V}_{i},\overline{X}_{2,i})\right]\nonumber\\
&= \sum_{i=0}^{n/D-1}   I_d(\overline{X}_{2,i};\overline{Y}_i|\tilde{V}_{i})
\end{flalign}
where (a) follows $\textbf{X}_1$ since is a function of $M_1$, (b) follows since $M_1,M_2-(\textbf{X}_1,\textbf{X}_2)-\textbf{Y}$ is a Markov chain for any given $d$,
(c) and (d) follow since conditioning reduces entropy and (e) follows since $(\textbf{X}_1,\textbf{X}_2,Y^{iD})-(\tilde{V}_{i},\overline{X}_{2,i},d)-\overline{Y}_i$ is a Markov chain for any \textsl{given} $d$ and all $i$.

Hence, we have that for every $d\in\mathcal{D}$
\begin{flalign}
n(R_1+R_2)&\leq \sum_{i=0}^{n/D-1}  I_d(\tilde{V}_{i},\overline{X}_{2,i};\overline{Y}_i)+n\delta_n\nonumber\\
nR_2&\leq \sum_{i=0}^{n/D-1}   I_d(\overline{X}_{2,i};\overline{Y}_i|\tilde{V}_{i})+n\delta_n'.
\end{flalign}
In addition, note that the delay $d$ is known only at the receiver, that is, $\textbf{X}_1$ and $\textbf{X}_2$ are not functions of the delay.

Let $Q$ be a random variable which is distributed uniformly over  $\{0,\ldots,n/D-1\}$ and independent of $\textbf{X}_1,\textbf{V},\textbf{X}_2,\textbf{Y}$, and let
 $\overline{X}_1=\overline{X}_{1,Q}$, $\tilde{V}=\tilde{V}_{Q}$,  $\overline{X}_2=\overline{X}_{2,Q}$ and $\overline{Y}=\overline{Y}_{Q}$. Then,
\begin{flalign}
R_1+R_2&\leq\frac{1}{n}\sum_{i=0}^{n/D-1}  I_d(\overline{X}_{1,i},\overline{X}_{2,i};\overline{Y}_i)+\delta_n \nonumber\\
&=\frac{1}{D}\cdot\frac{D}{n}\sum_{i=0}^{n/D-1}  I_d(\overline{X}_{1,i},\overline{X}_{2,i};\overline{Y}_i)+\delta_n \nonumber\\
&=\frac{1}{D}\cdot I_d(\overline{X}_1,\overline{X}_2;\overline{Y}|Q)+\delta_n
\end{flalign}

Similarly,
\begin{flalign}
R_2&\leq \frac{1}{n}\sum_{i=0}^{n/D-1}  I_d(\overline{X}_{2,i};\overline{Y}_i|\tilde{V}_{i})+\delta_n'\nonumber\\
&=  \frac{1}{D}\cdot\frac{D}{n}\sum_{i=0}^{n/D-1}  I_d(\overline{X}_{2,i};\overline{Y}_i|\tilde{V}_{i})+\delta_n' \nonumber\\
&=  \frac{1}{D}\cdot I_d(\overline{X}_2;\overline{Y}|\tilde{V},Q)+\delta_n'.
\end{flalign}

Since both encoders do not know the delay $d$ in advance, $X_1$ and $X_2$ do not depend on the delay $d$.
Therefore, after taking $n\rightarrow\infty$, we can write the following outer bound of the rate-region as aa function of $P(q),P(\tilde{v}|q),P(\overline{x}_2|\tilde{v},q)$, and $d\in\mathcal{D}$,
\begin{flalign}\label{MAC_upper}
&\overline{\mathcal{R}}_d(P(q),P(\tilde{v}|q),P(\overline{x}_2|\tilde{v},q))=
\left\{
\begin{array}{ll}
    \multirow{2}{*}{$(R_1,R_2):$}&R_1+R_2 \leq\frac{1}{D}\cdot I_d(\overline{X}_1,\overline{X}_2;\overline{Y}|Q)\\
&\hspace{1cm}R_2 \leq   \frac{1}{D}\cdot I_d(\overline{X}_2;\overline{Y}|\tilde{V},Q)
\end{array}
\right\}.
\end{flalign}

In addition, the fact that both encoders do not know in advance the delay $d$  means that a rate-pair is achievable only if it lies in the intersection of all the regions $\overline{R}_d(P(q),P(\tilde{v}|q),P(\overline{x}_2|\tilde{v},q))$. Denote
\begin{flalign}
\overline{\mathcal{R}}(P(q),P(\bar{\tilde{v}}|q),P(\overline{x}_2|\tilde{v},q))&=\bigcap_{d\in\mathcal{D}}
\overline{\mathcal{R}}_d(P(q),P(\bar{\tilde{v}}|q),P(\overline{x}_2|\tilde{v},q)).
\end{flalign}
Note that since the set of all possible delays is finite it follows that
\begin{flalign}
\overline{\mathcal{R}}(P(q),P(\overline{v}|q),P(\overline{x}_2|\tilde{v},q))&=\left\{
\begin{array}{ll}
    \multirow{2}{*}{$(R_1,R_2):$}&R_1+R_2 \leq \frac{1}{D}\cdot\min_{d\in\mathcal{D}} I_d(\overline{X}_1,\overline{X}_2;\overline{Y}|Q)\\
&\hspace{1cm}R_2 \leq\frac{1}{D}\cdot\min_{d\in\mathcal{D}} I_d(\overline{X}_2;\overline{Y}|\tilde{V},Q)
\end{array}\right\}.
\end{flalign}
The union over all p.m.f.'s $P(q),P(\tilde{v}|q)$, and $P(\overline{x}_2|\tilde{v},q)$ yields the outer rate-region
\begin{flalign}
\overline{\mathcal{R}}=\bigcup_{P(q),P(\tilde{v}|q),P(\overline{x}_2|\tilde{v},q)}\overline{\mathcal{R}}(P(q),P(\tilde{v}|q),P(\overline{x}_2|\tilde{v},q)).
\end{flalign}

\subsection{Proof of Lemma~\ref{lemma:lemma_Gaussian}}\label{ACMAC_Gaussian1}

 Recall the notations stated in (\ref{eq:dep_over_X1})-(\ref{eq:def_bar_vi}). This appendix includes the proof for Lemma \ref{lemma:lemma_Gaussian} which claims that it is sufficient to consider only Gaussian  random vector
\begin{flalign}
\overline{U}\triangleq(\overline{V}_{1,1},\overline{V}_{1,2},\overline{V}_{2,2},\overline{X}_{2,1},\overline{X}_{2,1},Q)^T
\end{flalign}
 for the outer bound (\ref{outer_region_ACMAC}). Moreover, it also suffices to consider a deterministic  $Q$.
We remark that we treat all the vectors in this appendix as column vectors. \begin{IEEEproof}

Our proof is composed of two parts, in the first part we find power constraints that permit
Theorem~\ref{ACMAC_outer_single} to be generalized to the Gaussian ACMAC, using standard techniques \cite{Gallager1968}. Then, we proceed to prove the rest of the Lemma.

\textit{Part I:}

By the model definition,
\begin{flalign}\label{eq:orig_const}
\frac{1}{n}\sum_{i=1}^nX_{1,i}^2\leq P_1,\quad \frac{1}{n}\sum_{i=1}^nX_{2,i}^2\leq P_1.
\end{flalign}

We note that the conditions
\begin{flalign}\label{eq:exp_const}
E\left(\frac{1}{n}\sum_{i=1}^nX_{1,i}^2\right)\leq P_1,\quad E\left(\frac{1}{n}\sum_{i=1}^nX_{2,i}^2\right)\leq P_2
\end{flalign}
are less restrictive then the ones in (\ref{eq:orig_const}). Therefore, replacing (\ref{eq:orig_const}) with (\ref{eq:exp_const}) may only enlarge the outer region.

Choosing the first $D$ symbols and the last $D$ symbol to be  zero , we conclude that  $\frac{1}{n}\sum_{i=1}^nX_{1,i-d}^2\leq P_1$ for every $d\in\mathcal{D}$.   Clearly, since $D$ is finite, this choice does not affect the outer-region.
Now, let $\overline{X}_{1,i}=X_{1,i D+1-d}^{(i+1) D-d}$, then
(see footnote~\ref{foot:expec_d}),\begin{flalign}
E\left(\frac{1}{n}\sum_{i=1}^nX_{1,i}^2\right)&=E_d\left(\frac{1}{n}\sum_{i=1}^nX_{1,i-d}^2\right)
\nonumber\\&=
\frac{1}{n}\sum_{i=0}^{n/D-1} E_d\left(\overline{X}_{1,i}^T\overline{X}_{1,i}\right)\nonumber\\
&=\frac{1}{D}\cdot\frac{D}{n}\sum_{i=0}^{n/D-1}E_d\left( \overline{X}_{1,i}^T\overline{X}_{1,i}\right).
\end{flalign}
Let $Q$ be, as before, a random variable  that is independent of $\textbf{X}_1$ and is distributed uniformly over the set $\{0,\ldots,n/D-1\}$ and let $\overline{X}_{1}\triangleq\overline{X}_{1,Q}$. Then
\begin{flalign}
E\left(\frac{1}{n}\sum_{i=1}^nX_{1,i}^2\right)
&=\frac{1}{D} \cdot E_d\left(\overline{X}_{1}^T\overline{X}_{1}|Q\right).
\end{flalign}

Similarly, we have that $\frac{1}{D} \cdot E\left(\overline{X}_{2}^T\overline{X}_{2}|Q\right)\leq P_2$.

Consequently, the union of all rate-pairs satisfying (\ref{outer_region_ACMAC}) for a joint distribution that satisfies (\ref{eq:outer_prob}) where $f(y|x_1,x_2)$  is defined by (\ref{Gassian_ACMAC_model}), under the constraints
\begin{flalign}\label{power_constraints}
&\frac{1}{2}E_d\left(\overline{X}_1^T \overline{X}_1|Q\right)\leq P_1,\quad \frac{1}{2}E\left(\overline{X}_2^T \overline{X}_2|Q\right)\leq P_2,\quad \forall d\in\mathcal{D}
\end{flalign}
includes the capacity region of the Gaussian ACMAC with $\mathcal{D}=\{0,1\}$.

\textit{Part II:}

Let $W_1$ and $W_2$ be column  random vectors, their covariance matrix is defined as
\begin{flalign}
\text{cov}(W_1,W_2)=E\left[(W_1-EW_1) (W_2-EW_2)^T \right].
\end{flalign}

Denote by  $C$ the covariance matrix of $\overline{U}$.
Further, let
\begin{flalign}
 C_{\overline{Y}}(d)&\triangleq\text{cov}_d(\overline{Y}),\nonumber\\
\sigma^2_Q&\triangleq  \text{var}(Q),\nonumber\\
C_{\overline{Y},Q}(d)&\triangleq  \text{cov}(\overline{Y},Q),\nonumber\\
C_{Q,\overline{Y}}(d)&\triangleq  \text{cov}(Q,\overline{Y}).
\end{flalign}

For each $d\in\mathcal{D}$ and $i\in\{1,2\}$, let\ $Y_i=X_{1,i+1-d}+X_{2,i}+Z_i$. Denote, $\overline{Z}=(Z_1,Z_2)^{T} $, then
\begin{flalign}
I_d(\overline{X}_1,\overline{X}_2;\overline{Y}|Q)&=h_d(\overline{Y}|Q)-h_d(\overline{Y}|\overline{X}_1,\overline{X}_2,Q)\nonumber\\
& = h_d(\overline{Y}|Q)-h_d(\overline{Z})\nonumber\\
&=h_d(\overline{Y}|Q)-\frac{1}{2}\log((2\pi e)^{2} N^2)\nonumber\\
&\stackrel{(*)}{\leq} \frac{1}{2}E_{Q}\log[\det(\text{cov}_d(\overline{Y}|Q))]-\frac{1}{2}\log(N^2)\nonumber\\
\end{flalign}
where $(*)$ follows from Lemma 1 in \cite{Thomas1987}, the equality is achieved whenever $\overline{U}$ is a Gaussian random vector.

Since $\bar{Z}$ is independent of $\overline{U}$,  and  the random vectors  $\overline{Z}$ and  $\overline{U}$ are jointly Gaussian, the covariance matrix
\begin{flalign}
\text{cov}_d(\overline{Y}|Q)=C_{\overline{Y}}(d)-C_{\overline{Y},Q}(d)&\sigma^{-2}_QC_{Q,\overline{Y}}(d)
\end{flalign}
 does not depend on the exact value of $Q$ but only on its distribution.
It follows that in that scenario
the expectation operator in the following line can be ignored as written below
\begin{flalign}\label{eq:gaussian_lemma1_notQ1}
I_d(\overline{X}_1,\overline{X}_2;\overline{Y}|Q)
&= \frac{1}{2}E_{Q}\log[\det(\text{cov}_d(\overline{Y}|Q))]-\frac{1}{2}\log(N^2)\nonumber\\
&\stackrel{}=\frac{1}{2} \log[\det(\text{cov}_d(\overline{Y}|Q))]-\frac{1}{2}\log(N^2).
\end{flalign}

We  proceed with analyzing the expression $I_d(\overline{X}_2;\overline{Y}|\overline{V},Q)$.
Let
\begin{flalign}\label{eq:def_U_check}
\check{U}=(\overline{V}_{1,1},\overline{V}_{1,2},\overline{V}_{2,2},Q)^{T},
\end{flalign}
 then
 \begin{flalign}
I_d(\overline{X}_2;\overline{Y}|\tilde{V},Q)&=I_d(\overline{X}_2;\overline{Y}|\check{U})\nonumber\\
&=h_d(\overline{Y}|\check{U})-h_d(\overline{Y}|\overline{X}_2,\check{U})\nonumber\\
&=h_d(\overline{Y}|\check{U})-\frac{1}{2}\log((2\pi e)^{2} N^2)\nonumber\\
&=h_d(\overline{X}_1+\overline{X}_2+\overline{Z}|\check{U})-\frac{1}{2}\log((2\pi e)^{2} N^2)\nonumber\\
&=h_d(\overline{X}_2+\overline{Z}|\check{U})-\frac{1}{2}\log((2\pi e)^{2} N^2)\nonumber\\
&\stackrel{(*)}{\leq}\frac{1}{2} E_{\tilde{X}_{1},Q}\log[(\text{cov}(\overline{X}_2+\overline{Z}|\check{U}))]-\frac{1}{2}\log(N^2)
\end{flalign}
where $(*)$ follows from Lemma 1 in \cite{Thomas1987} achieves the equality whenever $\overline{U}$ is a Gaussian random vector and $\overline{Z}$ is a Gaussian noise that is statistically independent of  $\overline{U}$.

Denote
\begin{flalign}\label{eq:cov_U_chaek_X2_Z}
C_{\overline{X}_2+\overline{Z}}&\triangleq\text{cov}(\overline{X}_2+\overline{Z})\nonumber\\
C_{\check{U},\overline{X}_2+\overline{Z}}&\triangleq\text{cov}_d(\check{U},\overline{X}_2)\nonumber\\
 C_{\overline{X}_2+\overline{Z},\check{U}}&\triangleq \text{cov}_d(\overline{X}_2+\overline{Z},\check{U})\nonumber\\
 C_{\check{U}}&\triangleq\text{cov}_d(\check{U}).
\end{flalign}
For jointly Gaussian $\overline{U}$ and $\overline{Z}$ the random vectors  $\check{U}$, $\overline{X}_2$  and  $\overline{Z}$ are jointly Gaussian and
the covariance matrix
\begin{flalign}
\text{cov}(\overline{X}_2+\overline{Z}|\check{U})&=C_{\overline{X}_2+\overline{Z}}-C_{\overline{X}_2+\overline{Z},\check{U}}C_{\check{U}} ^{-1}C_{\check{U},\overline{X}_2+\overline{Z}}\nonumber\\
\end{flalign}
 does not depend on the exact value of equivalently $\check{U}$ but only on their distribution. Let $\tilde{X}_1\triangleq(\overline{V}_{1,1},\overline{V}_{1,2},\overline{V}_{2,2})^T,$ it follows that for jointly Gaussian $\overline{U}$ and $\overline{Z}$
\begin{flalign}\label{eq:gaussian_lemma1_notQ2}
I_d(\overline{X}_2;\overline{Y}|\tilde{V},Q)
&=I_d(\overline{X}_2;\overline{Y}|\tilde{X}_1,Q)\nonumber\\
&=\frac{1}{2} E_{\tilde{X}_1,Q}\log[\det(\text{cov}(\overline{X}_2+\overline{Z}|\tilde{X}_1,Q))]-\frac{1}{2}\log(N^2)\nonumber\\
&{=}\frac{1}{2} \log[\det(\text{cov}(\overline{X}_2+\overline{Z}|\tilde{X}_1,Q))]-\frac{1}{2}\log(N^2).
\end{flalign}

By equations (\ref{eq:gaussian_lemma1_notQ1}) and (\ref{eq:gaussian_lemma1_notQ2}) it follows that for a Gaussian $\bar{U}$ the expressions $I_d(\overline{X}_1,\overline{X}_2;\overline{Y}|Q=q)$  are equal for every $q$; also, the expressions $I_d(\overline{X}_2;\overline{Y}|\tilde{V},Q=q)$ are equal for every $q$. Now, since $I_d(\overline{X}_1,\overline{X}_2;\overline{Y}|Q)\leq I_d(\overline{X}_1,\overline{X}_2;\overline{Y})$ and since $I_d(\overline{X}_2;\overline{Y}|\tilde{V},Q)\leq I_d(\overline{X}_2;\overline{Y}|\tilde{V})$ with equality if $Q$ is independent of $\overline{X}_2,\tilde{V}$ and $\overline{Y}$, it suffices to consider a deterministic  $Q$. We note that by the definitions of $\tilde{V}$ and $\tilde{X}_1$ the following equality holds
\begin{flalign}
I_d(\overline{X}_2;\overline{Y}|\tilde{V})=I_d(\overline{X}_2;\overline{Y}|\tilde{X}_{1}).
\end{flalign}
Further, since $Q$ is deterministic,
it follows from (\ref{power_constraints}) that the matrix $C$ must  fulfill the  conditions:
\begin{enumerate}
\item $C_{11}+C_{22}\leq 2P_1$,
\item $C_{22}+C_{33}\leq 2P_1$,
\item $C_{44}+C_{55}\leq 2P_2$.
\end{enumerate}
\end{IEEEproof}

\subsection{Proof of Proposition~\ref{props:Gauss_ACMAC_sub_coop}}\label{ACMAC_Gaussian2}
This appendix includes the proof of Proposition~\ref{props:Gauss_ACMAC_sub_coop}.
We remark that we treat to all the vectors in this appendix as column vectors.
\begin{IEEEproof}
Let $\mathcal{S}$ be the set of all real positive semi-definite matrices  such that
\begin{flalign}
\begin{pmatrix}
P_1 & 0 & 0 & C_{14} & C_{15}\\
0 & P_1 & 0 & C_{24} & C_{25}\\
0 & 0 & P_1 & C_{34} & C_{35}\\
C_{14} & C_{24} & C_{34} & C_{44} & C_{45} \\
C_{15} & C_{25} & C_{35} & C_{45} & C_{55}
\end{pmatrix}
\end{flalign}
and  $C_{44}+C_{55}\leq 2P_2$.

Recall that $\tilde{X}_1=(\overline{V}_{1,1},\overline{V}_{1,2},\overline{V}_{2,2})^{T}$ and denote $\overline{U}=(\overline{V}_{1,1},\overline{V}_{1,2},\overline{V}_{2,2},\bar{X}_{2,1},\bar{X}_{2,1})^{T}$ and assume that $\text{cov}(\overline{U})\triangleq C_{\overline{U}}\in\mathcal{S}$.
Also, let \begin{flalign}
 C_{\overline{Y}}(d) &\triangleq \text{cov}_d(\overline{Y})\nonumber\\
 C_{\overline{X}_2+\overline{Z}} &\triangleq\text{cov}(\overline{X}_2+\overline{Z})\nonumber\\
 C_{\tilde{X}_1,\overline{X}_2}&\triangleq \text{cov}(\tilde{X}_1,\overline{X}_2)\nonumber\\
  C_{\overline{X}_2,\tilde{X}_1}&\triangleq \text{cov}(\overline{X}_2,\tilde{X}_1)\nonumber\\
 C_{\tilde{X}_1} &\triangleq \text{cov}(\tilde{X}_1).
\end{flalign}

Since,
\begin{flalign}
\text{cov}(\overline{X}_2+\overline{Z}|\tilde{X}_1)=C_{\overline{X}_2+\overline{Z}}-C_{\overline{X}_2,\tilde{X}_1}C_{\tilde{X}_1}^{-1}C_{\tilde{X}_1,\overline{X}_2},
\end{flalign}
by the proof of Lemma \ref{lemma:lemma_Gaussian}  the region
\begin{flalign}\label{eq:outer_Gauss_region_ACMAC}
\tilde{\mathcal{R}}&=\bigcup_{C_{\overline{U}}\in\mathcal{S}}
\left\{
\begin{array}{ll}
    \multirow{2}{*}{\hspace{-0.2cm}$(R_1,R_2):$}\hspace{-0.3cm}&R_1+R_2 \leq\frac{1}{4}\min_{d\in\{0,1\}}\log\det[C_{\overline{Y}}(d)]-\frac{1}{2}\log N\\
&\hspace{1cm}R_2 \leq   \frac{1}{4}\min_{d\in\{0,1\}}\log\det[C_{\overline{X}_2+\overline{Z}}-C_{\overline{X}_2\tilde{X}_1}C_{\tilde{X}_1}^{-1}C_{\tilde{X}_1\overline{X}_2}]-\frac{1}{2}\log N
\end{array}\right\}
\end{flalign}
includes all the rate-pairs which are achieved  by covariance matrices in the set $\mathcal{S}$.

We now analyze that covariance matrices $C_{\tilde{X}_1}^{-1},C_{\overline{X}_2,\tilde{X}_1},C_{\overline{X}_2+\overline{Z}}$ and $C_{\overline{Y}}(d)$.
\begin{flalign}
C_{\tilde{X}_1}^{-1}&=P_1^{-1}I,\nonumber\\
C_{\overline{X}_2,\tilde{X}_1}&=\begin{pmatrix}
C_{14}& C_{24} & C_{34}\\
C_{15}& C_{25} & C_{35}
\end{pmatrix},\nonumber\\
C_{\overline{X}_2+\overline{Z}}&=C_{\overline{X}_2}+C_{\overline{Z}}=\begin{pmatrix}
C_{44} & C_{45}\\
C_{45} & C_{55}
\end{pmatrix}+NI=\begin{pmatrix}
C_{44}+N & C_{45}\\
C_{45} & C_{55}+N
\end{pmatrix} \nonumber\\
C_{\overline{Y}}(d=0)&=  \begin{pmatrix}
P_{1} & 0\\
0 & P_{1}
\end{pmatrix} +\begin{pmatrix}
2C_{24} & C_{34}+C_{25}\\
C_{34}+C_{25}& 2C_{35}
\end{pmatrix}+\begin{pmatrix}
C_{44} & C_{45}\\
C_{45} & C_{55}
\end{pmatrix}+NI\nonumber\\
&=  \begin{pmatrix}
P_{1}+2C_{24}+C_{44}+N & C_{34}+C_{25}+C_{45}\\
 C_{34}+C_{25}+C_{45} & P_{1}+2C_{35}+C_{55}+N
\end{pmatrix} \nonumber\\
C_{\overline{Y}}(d=1)&=  \begin{pmatrix}
P_{1} & 0\\
0 & P_{1}
\end{pmatrix} +\begin{pmatrix}
2C_{14} & C_{24}+C_{15}\\
C_{24}+C_{15}& 2C_{25}
\end{pmatrix}+\begin{pmatrix}
C_{44} & C_{45}\\
C_{45} & C_{55}
\end{pmatrix}+NI\nonumber\\
&=  \begin{pmatrix}
P_{1}+2C_{14}+C_{44}+N & C_{24}+C_{15}+C_{45}\\
 C_{24}+C_{15}+C_{45} & C_{22}+2C_{25}+C_{55}+N
\end{pmatrix}.
\end{flalign}
Additionally,
\begin{flalign}
C_{\overline{X}_2+\overline{Z}}-C_{\overline{X}_2\tilde{X}_1}C_{\tilde{X}_1}^{-1}C_{\tilde{X}_1\overline{X}_2}
&=\begin{pmatrix}
C_{44}+N & C_{45}\\
C_{45} & C_{55}+N
\end{pmatrix}-\frac{1}{P_1}\begin{pmatrix}
\sum_{i=1}^3 C_{i4}^2 & \sum_{i=1}^3 C_{i4}C_{i5}\\
\sum_{i=1}^3 C_{i4}C_{i5} & \sum_{i=1}^3 C_{i5}^2
\end{pmatrix}\nonumber\\
&=\begin{pmatrix}
C_{44}+N-\frac{1}{P_1}\sum_{i=1}^3 C_{i4}^2 & C_{45}-\frac{1}{P_1}\sum_{i=1}^3 C_{i4}C_{i5}\\
C_{45}-\frac{1}{P_1}\sum_{i=1}^3 C_{i4}C_{i5} & C_{55}+N-\frac{1}{P_1}\sum_{i=1}^3 C_{i5}^2
\end{pmatrix}
\end{flalign}
It follows that for each covariance matrix $C_{\overline{U}}\in\mathcal{S}$
\begin{flalign}\label{eq:upper_R12a}
&R_1+R_2\leq\frac{1}{4}\min_{d\in\{0,1\}}\log\det[C_{\overline{Y}}(d)]-\frac{1}{2}\log N \nonumber\\
&\leq \frac{1}{4} \min\left\{\log(P_{1}+2C_{24}+C_{44}+N)+\log(P_{1}+2C_{35}+C_{55}+N),\right.\nonumber\\
&\hspace{1.8cm}\left. \log(P_{1}+2C_{14}+C_{44}+N)+\log(P_{1}+2C_{25}+C_{55}+N)\right\}-\frac{1}{2}\log N\nonumber\\
&= \frac{1}{4} \min\left\{\log(P_{1}+2\rho_{24}\sqrt{P_1C_{44}}+C_{44}+N)+\log(P_{1}+2\rho_{35}\sqrt{P_1C_{55}}+C_{55}+N),\right.\nonumber\\
&\hspace{1.8cm}\left. \log(P_{1}+2\rho_{14}\sqrt{P_1C_{44}}+C_{44}+N)+\log(P_{1}+2\rho_{25}\sqrt{P_1C_{55}}+C_{55}+N)\right\}-\frac{1}{2}\log N
\end{flalign}
and that
\begin{flalign}\label{eq:upper_R2a}
R_2&  \leq  \frac{1}{4}\min_{d\in\{0,1\}}\log\det[C_{\overline{X}_2+\overline{Z}}-C_{\overline{X}_2,\tilde{X}_1}C_{\tilde{X}_1}^{-1}C_{\tilde{X}_1,\overline{X}_2}]-\frac{1}{2}\log N \nonumber\\
&\leq \frac{1}{4}\left[\log\left(C_{44}+N-\frac{1}{P_1}\sum_{i=1}^3 C_{i4}^2\right)+\log\left(C_{55}+N-\frac{1}{P_1}\sum_{i=1}^3 C_{i5}^2\right)\right]-\frac{1}{2}\log N\nonumber\\
&= \frac{1}{4}\left[\log\left(C_{44}(1-\rho_{14}^2-\rho_{24}^2-\rho_{34}^2)+N\right)+\log\left(C_{55}(1-\rho_{15}^2-\rho_{25}^2-\rho_{35}^2)+N\right)\right]-\frac{1}{2}\log N.
\end{flalign}
Since the terms $\rho_{34}$ and  $\rho_{15}$ do not appear in (\ref{eq:upper_R12a}) we can choose their optimal value for  (\ref{eq:upper_R2a}), that is, $\rho_{34}=\rho_{15}=0$. Choosing these values yields the following bound
\begin{flalign}\label{eq:upper_R2b}
R_2&  \leq   \frac{1}{4}\left[\log\left(C_{44}(1-\rho_{14}^2-\rho_{24}^2)+N\right)+\log\left(C_{55}(1-\rho_{25}^2-\rho_{35}^2)+N\right)\right]-\frac{1}{2}\log N.
\end{flalign}
In addition,   considering only  $\rho_{14},\rho_{24},\rho_{25},\rho_{35}\geq 0$ does not affect  (\ref{eq:upper_R2b}) while it can only increase (\ref{eq:upper_R12a}), therefore, hereafter we only consider nonnegative $\rho_{14},\rho_{24},\rho_{25}$ and $\rho_{35}$.

Since $C_{\overline{U}}$ is a covariance matrix, by definition it is positive semidefinite, therefore, $1\geq\rho_{14}^2+\rho_{24}^2$ and $1\geq\rho_{25}^2+\rho_{35}^2$.
As stated above it is sufficient to consider for (\ref{eq:upper_R12a}) and (\ref{eq:upper_R2b}) only $\rho_{14},\rho_{24},\rho_{25},\rho_{35}\geq 0$. Combining these two facts, we infer that the optimal choice for  (\ref{eq:upper_R12a}) and (\ref{eq:upper_R2b}) holds that  $C_{44}+C_{55}=2P_2$. Additionally, by the concavity of the logarithmic function,
\begin{flalign}
&\frac{1}{4}\left[\log(P_{1}+2\rho_{24}\sqrt{P_1C_{44}}+C_{44}+N)+\log(P_{1}+2\rho_{35}\sqrt{P_1(2P_2-C_{44})}+2P_2-C_{44}+N)\right]\nonumber\\
&\leq \frac{1}{2}\log \left(P_{1}+\sqrt{P_1}\left(\rho_{24}\sqrt{C_{44}}+\rho_{35}\sqrt{2P_2-C_{44}}\right)+P_2+N\right)\nonumber\\
&\frac{1}{4}\left[\log(P_{1}+2\rho_{14}\sqrt{P_1C_{44}}+C_{44}+N)+\log(P_{1}+2\rho_{25}\sqrt{P_1(2P_2-C_{44})}+2P_2-C_{44}+N)\right]\nonumber\\
&\leq \frac{1}{2}\log \left(P_{1}+\sqrt{P_1}\left(\rho_{14}\sqrt{C_{44}}+\rho_{25}\sqrt{2P_2-C_{44}}\right)+P_2+N\right)
\end{flalign}
and
\begin{flalign}
&\frac{1}{4}\left[\log\left(C_{44}(1-\rho_{14}^2-\rho_{24}^2)+N\right)+\log\left(C_{55}(1-\rho_{25}^2-\rho_{35}^2)+N\right)\right]\nonumber\\
&\leq \frac{1}{2}\log\left(\frac{1}{2}C_{44}(1-\rho_{14}^2-\rho_{24}^2)+\frac{1}{2}(2P_2-C_{44})(1-\rho_{25}^2-\rho_{35}^2)+N\right).
\end{flalign}
This leads to the following conclusion. Let
\begin{flalign}
\mathcal{K}=\{(\rho_{14},\rho_{24},\rho_{25},\rho_{35},C_{44})\in\mathbb{R}^5|\:&\rho_{14},\rho_{24},\rho_{25},\rho_{35},C_{44}\geq0,\nonumber\\
&\rho_{14}^2+\rho_{24}^2\leq1,\nonumber\\
&\rho_{25}^2+\rho_{35}^2\leq1,\nonumber\\
& 0\leq C_{44} \leq 2P_2\}
\end{flalign}
and let
\begin{flalign}
R_{12}^0 (\rho_{24},\rho_{35},C_{44})&\triangleq\frac{1}{2}\log \left[\frac{1}{N}\left(P_{1}+\sqrt{P_1}\left(\rho_{24}\sqrt{C_{44}}+\rho_{35}\sqrt{2P_2-C_{44}}\right)+P_2+N\right)\right]\nonumber\\
R_{12}^1 (\rho_{14},\rho_{25},C_{44})&\triangleq\frac{1}{2}\log \left[\frac{1}{N}\left(P_{1}+\sqrt{P_1}\left(\rho_{14}\sqrt{C_{44}}+\rho_{25}\sqrt{2P_2-C_{44}}\right)+P_2+N\right)\right]\nonumber\\
R_2  (\rho_{14},\rho_{24},\rho_{25},\rho_{35},C_{44})&\triangleq \frac{1}{2}\log\left[\frac{1}{N}\left(\frac{C_{44}}{2}(1-\rho_{14}^2-\rho_{24}^2)+\frac{(2P_2-C_{44})}{2}(1-\rho_{25}^2-\rho_{35}^2)+N\right)\right].
\end{flalign}

 The rate-region which is composed of all the rate-pairs $(R_1,R_2)$ such that
\begin{flalign}\label{eq:R12_upper2}
R_1+R_2&\leq\min\left\{
R_{12}^0 (\rho_{24},\rho_{35},C_{44}),R_{12}^1 (\rho_{14},\rho_{25},C_{44})\right\}
\end{flalign}
and
\begin{flalign}\label{eq:R2_upper2}
R_2 \leq  R_2  (\rho_{14},\rho_{24},\rho_{25},\rho_{35},C_{44})
\end{flalign}
for some $(\rho_{14},\rho_{24},\rho_{25},\rho_{35},C_{44})\in\mathcal{K}$ includes the rate-region $\tilde{\mathcal{R}}$ (see (\ref{eq:outer_Gauss_region_ACMAC})).

We next prove that it suffices to  consider $\rho_{14}=\rho_{24}$ and $\rho_{25}=\rho_{35}$.  We prove this property by proving that for every $\rho_{14},\rho_{24},\rho_{25},\rho_{35}$ there exist $\rho_{14}^*=\rho_{24}^*$ and $\rho_{25}^*=\rho_{35}^*$ which increase both (\ref{eq:R12_upper2}) and (\ref{eq:R2_upper2}). Choose $\rho_{14}^*=\rho_{24}^*=\frac{1}{2}(\rho_{14}+\rho_{24})$ and $\rho_{25}^*=\rho_{35}^*=\frac{1}{2}(\rho_{25}+\rho_{35})$ and\footnote{We note that for every $(\rho_{14},\rho_{24},\rho_{25},\rho_{35})$ which fulfills the conditions of the  region $\mathcal{K}$, the  $4$-tuple $(\rho_{14}^*,\rho_{24}^*,\rho_{25}^*,\rho_{35}^*)$ also fulfills the conditions of the region $\mathcal{K}$.} assume without loss of generality that $R_{12}^0 (\rho_{24},\rho_{15},C_{44})\leq R_{12}^1 (\rho_{14},\rho_{25},C_{44})$.
Then,
\begin{flalign}\label{eq:indX1p2_begin}
&(1-\rho_{14}^2-\rho_{24}^2)-(1-(\rho_{14}^*)^2-(\rho_{24}^*)^2)=(\rho_{14}^*)^2+(\rho_{24}^*)^2-\rho_{14}^2-\rho_{24}^2\nonumber\\
&=\frac{1}{2}(\rho_{14}+\rho_{24})^{2}-\rho_{14}^2-\rho_{24}^{2}=-\frac{1}{2}(\rho_{14}-\rho_{24})^2\leq0.
\end{flalign}
Similarly, $(1-\rho_{25}^2-\rho_{35}^2)-(1-(\rho_{25}^*)^2-(\rho_{35}^*)^2) \leq 0$. Thus $R_2  (\rho_{14},\rho_{24},\rho_{25},\rho_{35},C_{44})\leq R_2  (\rho_{14}^*,\rho_{24}^*,\rho_{25}^*,\rho_{35}^*,C_{44})$.

Moreover, if $R_{12}^0 (\rho_{24},\rho_{15},C_{44})\leq R_{12}^1 (\rho_{14},\rho_{25},C_{44})$ then
\begin{flalign}
\rho_{24}\sqrt{C_{44}}+\rho_{35}\sqrt{2P_2-C_{44}}\leq\rho_{14}\sqrt{C_{44}}+\rho_{25}\sqrt{2P_2-C_{44}}.
\end{flalign}
It follows that
\begin{flalign}
\rho_{24}^*\sqrt{C_{44}}+\rho_{35}^*\sqrt{2P_2-C_{44}}&=\frac{1}{2}(\rho_{24}+\rho_{24})\sqrt{C_{44}}+\frac{1}{2}(\rho_{25}+\rho_{35})\sqrt{2P_2-C_{44}}\nonumber\\
&\geq\rho_{24}\sqrt{C_{44}}+\rho_{35}\sqrt{2P_2-C_{44}}.
\end{flalign}
Consequently,
\begin{flalign}\label{eq:indX1p2_end}
\min\{R_{12}^0 (\rho_{24},\rho_{15},C_{44}), R_{12}^1 (\rho_{14},\rho_{25},C_{44})\}\leq R_{12}^0 (\rho_{24}^*,\rho_{15}^*,C_{44})
\end{flalign}
for all $(\rho_{14},\rho_{24},\rho_{25},\rho_{35},C_{44})\in\mathcal{K}$.
By Equations (\ref{eq:indX1p2_begin})-(\ref{eq:indX1p2_end}) we  conclude that it is optimal to choose $\rho_{14}=\rho_{24}$ and $\rho_{25}=\rho_{35}$. Let
\begin{flalign}
\mathcal{K}_3\triangleq \{(\rho_{14},\rho_{25},C_{44})\in\mathbb{R}^3|\:&\rho_{14},\rho_{25},C_{44}\geq0,\nonumber\\
&2\rho_{14}^2\leq1,\nonumber\\
&\rho_{25}^2\leq1,\nonumber\\
& 0\leq C_{44} \leq 2P_2\},
\end{flalign} the above discussion the following outer-region 
\begin{flalign}\label{outer_Gauss_region_ACMAC_ind_temp}
\hat{\mathcal{R}}&=\bigcup_{\mathcal{K}_3}
\left\{
\begin{array}{ll}
    \multirow{2}{*}{\hspace{-0.2cm}$(R_1,R_2):$}\hspace{-0.3cm}&R_1+R_2 \leq\frac{1}{2}\log \left[\frac{1}{N}\left(P_{1}+\sqrt{P_1}\left(\rho_{14}\sqrt{C_{44}}+\rho_{25}\sqrt{2P_2-C_{44}}\right)+P_2+N\right)\right]\\
&\hspace{1cm}R_2 \leq   \frac{1}{2}\log\left[\frac{1}{N}\left(\frac{1}{2}C_{44}(1-2\rho_{14}^2)+\frac{1}{2}(2P_2-C_{44})(1-2\rho_{25}^2)+N\right)\right]
\end{array}\right\}.
\end{flalign}
We next prove that it suffices to consider $C_{44}=P_2$ and $\rho_{14}=\rho_{25}$. This is true since for each $(\rho_{14},\rho_{25},C_{44})\in\mathcal{K}_3$ we can find $\rho\in[0,1/\sqrt{2}]$ such that
  \begin{flalign}
\rho_{14}\sqrt{C_{44}}+\rho_{25}\sqrt{2P_2-C_{44}}&\leq 2\rho\sqrt{P_2}\nonumber\\
\frac{1}{2}C_{44}(1-2\rho_{14}^2)+\frac{1}{2}(2P_2-C_{44})(1-2\rho_{25}^2)&\leq P_2(1-2\rho^2).
 \end{flalign}
A simple algebra yields the following nonempty region
 \begin{flalign}\label{eq:rho_region}
 \frac{\rho_{14}}{2}\sqrt{\frac{C_{44}}{P_2}}+\frac{\rho_{25}}{2}\sqrt{2-\frac{C_{44}}{P_2}}&\leq \rho \leq \sqrt{\frac{C_{44}}{P_2}\cdot\frac{\rho_{14}^2}{2}+\left(2-\frac{C_{44}}{P_2}\right)\frac{\rho_{25}^2}{2}}.
 \end{flalign}
 Note that for $(\rho_{14},\rho_{25},C_{44})\in\mathcal{K}_3$ the region ($\ref{eq:rho_region}$) is included in $[0,1/\sqrt{2}]$; this concludes the proof.
\end{IEEEproof}

\subsection{Proof of Theorem~\ref{ACC-MAC_capacity}}\label{ACCMAC_C1}

The proof of Theorem~\ref{ACC-MAC_capacity} is similar to the proof of Theorem \ref{ACMAC_capacity}, for the sake of brevity we only outline the proof.
By sending predefined training sequences in the first $o(n)$ bits, the decoder can deduce the delay with probability of error that vanishes as $n$ tends to infinity. Therefore, we can assume that the decoder knows the delay $d$. The remaining of the  coding scheme can be described in the following manner. Partition each of the input sequences to sequences of length $k$ (hyper-symbols). Generate the codebooks of the two encoders using superposition coding according to p.m.f.'s of the hyper symbols. That is, for each of the codewords in the uninformed encoder's codebook the informed encoder generates a separate codebook. Let $C^{(1)}$ be the codebook of the uninformed encoder, and let $C^{(1)}(m)$ be the $m$-th codeword in $C^{(1)}$.  Denote by $C^{(2)}(m)$ the codebook of the informed encoder which is associated with the codeword $C^{(1)}(m)$. Further, the $k$-th codeword in $C^{(2)}(m)$ is denoted by $C^{(2)}(m,k)$.
Suppose that the uninformed and the informed encoders transmit the messages $m$ and $k$, respectively. In the encoding stage, the uninformed encoder transmits $C^{(1)}(m)$ to the informed encoder. \textit{The informed encoder first verifies that the codeword to be transmitted by the uninformed encoder is unique}\footnote{We note that this stage is the part in which the proofs of Theorems \ref{ACMAC_capacity} and \ref{ACC-MAC_capacity} differ.} (\textit{does not appear more than once in $C^{(1)}$)}. Then, the uninformed and the informed encoders transmit $C^{(1)}(m)$ and $C^{(2)}(m,k)$, respectively. Finally, when the decoder receives the output sequence, it partitions it to sequences of length $k$, and then discards the first $d_{max}$ and the last $d_{min}$ symbols of every hyper symbol. This process yields a modified output sequence whose hyper-symbols are statistically independent given the input hyper-symbols. Therefore, standard typicality techniques can be used to prove that the resulting average probability of error vanishes as $n$ tends to infinity.

\subsection{Proof of Theorem~\ref{ACC-MAC_inner_single}}\label{ACCMAC_A2}

\textbf{Codebook Generation:}
The codebooks $\mathcal{C}^{(1)},\mathcal{C}^{(1')}$ and $\mathcal{C}^{(2)}(l), 1 \leq l\leq |\mathcal{C}^{(1)}|$ are produced in the following manner:
Set $P_{X_1}(x_1),P_{X_2|\overline{V}}(x_2|\overline{v})$. Let $\mathcal{C}^{(1)}$ be the codebook of the common message $M_1$, which consists of $2^{nR_1}$ codewords, each of these codewords  is generated  according to $P(\textbf{x}_1)=\prod_{i=1}^n P_{X_1}(x_{1,i})$.

The codebook $\mathcal{C}^{(1')}$ is produced from $\mathcal{C}^{(1)}$ by the following one-to-one mapping: let $\textbf{x}_1(l)$ be the $l$th
codeword in  $\mathcal{C}^{(1)}$, then for every $i\in \{1,\ldots,n\}$ define
\begin{flalign}
\overline{v}_i(l)=(x_{1,i-d_{max}}(l),\ldots,x_{1,i+d_{min}}(l)).
\end{flalign}
The resulting codeword $\textbf{v}(l)=(\overline{v}_1(l),\ldots,\overline{v}_n(l))$ is the $l$-th codeword in $C^{(1')}$,
that is, the codewords in $\mathcal{C}^{(1)}$ appear in $\mathcal{C}^{(1')}$ as vectors that were produced by a sliding window of size $D$ on the sequence $\textbf{x}_1$.

Now, for every $\textbf{v}(l)\in\mathcal{C}^{(1')}$ generate randomly and independently $2^{nR_2}$
codewords\newline  $\{\textbf{x}_2(l,1),\ldots,\textbf{x}_2(l,2^{nR_2})\}$ according to
$P(\textbf{x}_2|\textbf{v})=\prod_{i=1}^n P_{X_2|\overline{V}}(x_{2,i}|\overline{v}_i)$.

 We denote
$\{\textbf{x}_2(l,1),\ldots,\textbf{x}_2(l,2^{nR_2})\}$ by $\mathcal{C}^{(2)}(l)$.

\textbf{Encoding:}
To send the messages $m_1,m_2$ encoder 1 sends $\textbf{x}_1(m_1)$.
Encoder 2 checks if $\textbf{x}_1(m_1)$ is unique in $\mathcal{C}^{(1)}$, that is, if there is a unique $\hat{m}_1\in \{1,\ldots,2^{nR}\}$
such that $\textbf{x}_1(\hat{m}_1)=\textbf{x}_1(m_1)$. If there is, encoder 2 sends  $\textbf{x}_2(\hat{m}_1,m_2)$, otherwise an error is declared and encoder 2 sends a sequence of zeroes.

\textbf{Decoding:}
Suppose that the actual delay in the channel is $d\in\mathcal{D}$, and
let \begin{flalign}
P_d(\overline{v},x_1,x_2,y)=P(\overline{v})\mathbbm{1}_{\{x_1=v_{d_{max}-d+1}\}}P(x_2|\overline{v})P(y|x_1,x_2),
\end{flalign}
and
 $T_{d,\epsilon}^n(\overline{V},X_2,Y)$ be the set of all vectors $(\textbf{v},\textbf{x}_2,\textbf{y})$ that are $\epsilon$-strongly typical with respect to $P_d(\overline{v},x_2,y)$.
 Given that the decoder knows the delay $d$, it  looks for a pair of messages $(\hat{m}_1,\hat{m}_2)\in\{1,\ldots,2^{nR_1}\}\times \{1,\ldots,2^{nR_2}\}$ such that
\begin{flalign}
(\textbf{v}(\hat{m}_1),\textbf{x}_2(\hat{m}_1,\hat{m}_2),\textbf{y})\in T_{d,\epsilon}^n(\overline{V},X_2,Y).
\end{flalign}
If there is no such pair or if there is more then one, an error is declared.

\textbf{Analysis of the Probability of Error:}
Suppose that the pair of messages $(m_1,m_2)=(1,1)$ is sent, and that the delay is $d\in\mathcal{D}$. An error is made if one or more of the following events occur:
\begin{flalign}
&\mathcal{E}_0=\{\textbf{x}_1(1)\notin T_{\epsilon}^n(X_1)\}\nonumber\\
&\mathcal{E}_1=\{\exists \hat{m}_1\neq 1 \text{ s.t. }\textbf{x}_1(\hat{m}_1)=\textbf{x}_1(1)\}\nonumber\\
&\mathcal{E}_2=\{\textbf{v}(1)\notin T_{\epsilon}^n(\overline{V})\}\nonumber\\
&\mathcal{E}_3=\{(\textbf{v}(1),\textbf{x}_2(1,1))\notin T_{\epsilon}^n(\overline{V},X_2)\}\nonumber\\
&\mathcal{E}_4=\begin{Bmatrix}(\textbf{v}(1),\textbf{x}_2(1,1),\textbf{y})\notin T_{d,\epsilon}^n(\overline{V},X_2,Y)\end{Bmatrix}\nonumber\\
&\mathcal{E}_5=\begin{Bmatrix}\exists \hat{m}_1\neq1 \text { and } \hat{m}_2\in\{1,\ldots,2^{nR_2}\} \text{ s.t. }
(\textbf{v}(\hat{m}_1),\textbf{x}_2(\hat{m}_1,\hat{m}_2),\textbf{y})\in T_{d,\epsilon}^n(\overline{V},X_2,Y) \end{Bmatrix}\nonumber\\
&\mathcal{E}_6=\begin{Bmatrix}\exists  \hat{m}_2\in\{1,\ldots,2^{nR_2}\} \text{ s.t. }
(\textbf{v}(1),\textbf{x}_2(1,\hat{m}_2),\textbf{y})\in T_{d,\epsilon}^n(\overline{V},X_2,Y). \end{Bmatrix}
\end{flalign}

By the union bound,
\begin{flalign}
\Pr(\mathcal{E})=\Pr\left(\bigcup_{i=0}^6 \mathcal{E}_i\right)\leq \Pr(\mathcal{E}_0)
&+\Pr(\mathcal{E}_1\cap \mathcal{E}_0^c)+\Pr(\mathcal{E}_1^c\cap \mathcal{E}_2)\nonumber\\
&+\Pr(\mathcal{E}_2^c\cap \mathcal{E}_3) + \Pr(\mathcal{E}_3^c\cap \mathcal{E}_4)+\Pr(\mathcal{E}_5\cap\mathcal{E}_1^c)+\Pr(\mathcal{E}_6)
\end{flalign}

From the LLN $\Pr(\mathcal{E}_0)\rightarrow0$ as $n\rightarrow\infty$. Additionally,
every sequence in the codebook of encoder 1 is generated by a memoryless source with p.m.f. $P_{X_1}$,
therefore   $\Pr(\mathcal{E}_1\cap \mathcal{E}_0^c)\leq e^{n\tilde{R}_1}2^{-nH(\tilde{X}_1)}$.
Thats is, $\Pr(\mathcal{E}_1\cap \mathcal{E}_0^c)\rightarrow 0$ as $\tilde{n}\rightarrow\infty$ if
$R_1<H(X_1)$.
Further, from the stationarity and ergodicity of $v^{n}$ we infer that $\Pr(\mathcal{E}_2)\rightarrow 0$ as $n\rightarrow\infty$.

By the conditional typicality lemma  \cite[p. 27]{NetworkInformationTheory}, $\Pr(\mathcal{E}_3)\rightarrow 0$ and $\Pr(\mathcal{E}_4)\rightarrow 0$ as $n\rightarrow\infty$.

We now bound $\Pr(\mathcal{E}_5,\textbf{y})$. Since both $(\textbf{v}(m_1),\textbf{x}_2(m_1,m_2))$ and $\textbf{y}$ are not generated according to an i.i.d. distribution, we cannot use the packing lemma. We therefore upper bound the probability that the output of the channel $\textbf{y}$ is accidently typical with the pair of typical sequences of the form $(\textbf{v},\textbf{x}_2)$. We denote this probability by $\Pr(\mathcal{E}_5,\textbf{y})$.
Let
\begin{flalign}
A(\textbf{y})=\begin{Bmatrix}
(\textbf{x}_1,\textbf{x}_2)\in \mathcal{X}_1^n\times \mathcal{X}_2^n:&(\textbf{v},\textbf{x}_2,\textbf{y})\in T_{d,\epsilon}^n(\overline{V},X_2,Y)
\end{Bmatrix}.
\end{flalign}
By definition of $\textbf{v}$, it follows that
\begin{flalign}
A(\textbf{y})=
\left\{
\begin{array}{ll}
 \multirow{2}{*}{$(\textbf{x}_1,\textbf{x}_2)\in \mathcal{X}_1^n\times \mathcal{X}_2^n:$}&\hspace{1cm}(\textbf{v},\textbf{x}_2,\textbf{y})\in T_{d,\epsilon}^n(\overline{V},X_2,Y)\\
&(\sigma(\textbf{x}_1(1),d),\textbf{x}_2(1,1),\textbf{y})\in T_{d,\epsilon}^n(X_1,X_2,Y)
\end{array}
\right\}
\end{flalign}
where the function $\sigma(\cdot,\cdot)$ is defined in (\ref{sigma_def}).

Further,
\begin{flalign}
\Pr(\mathcal{E}_5,\textbf{y})=\sum_{(\textbf{x}_1,\textbf{x}_2)\in A(\textbf{y})} P(\textbf{x}_1,\textbf{x}_2)=\sum_{(\textbf{x}_1,\textbf{x}_2)\in A(\textbf{y})} P(\textbf{x}_1)P(\textbf{x}_2|\textbf{x}_1).
\end{flalign}
Now, we the sequence $\textbf{x}_1$ according to an i.i.d. distribution, that is, $P(\textbf{x}_1)=\prod_{i=1}^n P(x_{1,i})$. Additionally,
\begin{flalign}
P(\textbf{x}_2|\textbf{x}_1)=\prod_{i=1}^n P(x_{2,i}|\overline{v}_i)
\end{flalign}
where $\overline{v}_i=(x_{1,i-d_{max}},\ldots,x_{1,i+d_{min}})$.
Therefore,
\begin{flalign}
\Pr(\mathcal{E}_5,\textbf{y})&=\sum_{(\textbf{x}_1,\textbf{x}_2)\in A(\textbf{y})} \prod_{i=1}^n P(x_{1,i})\prod_{i=1}^nP(x_{2,i}|\overline{v}_i)
\end{flalign}
Now, since $\textbf{x}_1$ is typical there exists $\epsilon_1(\delta)$ such that $\epsilon_1(\delta)\rightarrow 0$ as $\delta\rightarrow 0$, and
\begin{flalign}
\prod_{i=1}^n P(x_{1,i})\leq 2^{-n[H(X_1)-\epsilon_1(\delta)]}.
\end{flalign}
In addition, since $(\tilde{\textbf{v}},\textbf{x}_2)$ are jointly typical, there exists $\epsilon_2(\delta)$ such that $\epsilon_2(\delta)\rightarrow 0$ as $\delta\rightarrow 0$, and
\begin{flalign}
\prod_{i=1}^n P(x_{2,i}|\overline{v}_i)\leq 2^{-n[H(X_2|\overline{V})-\epsilon_2(\delta)]}.
\end{flalign}

We have that
\begin{flalign}
\Pr(\mathcal{E}_5,\textbf{y})&\leq \sum_{(\textbf{x}_1,\textbf{x}_2)\in A}
2^{-n[H(X_1)-\epsilon_1(\delta)]}
2^{-n[H(X_2|\overline{V})-\epsilon_2(\delta)]}.\\
\end{flalign}

By typicality, there are no more than $2^{H_d(X_1|Y)+\epsilon_3(n)}$ sequences $\textbf{x}_1$  in the set $A(\textbf{y})$, where $\epsilon_3(n)$ vanishes as $n$ tends to infinity. Additionally, for each $\textbf{x}_1$ there are at most $2^{H_d(X_2|Y,\overline{V})+\epsilon_4(n)}$ sequences $\textbf{x}_2$  in the set $A(\textbf{y})$, where $\epsilon_4(n)$ vanishes as $n$ tends to infinity. Therefore,
\begin{flalign}
\Pr(\mathcal{E}_5,\textbf{y})& \leq
2^{H_d(X_1|Y)+\epsilon_3(n)+H_d(X_2|Y,\overline{V})+\epsilon_4(n)}
2^{-n[H(X_1)-\epsilon_1(\delta)+H(X_2|\overline{V})-\epsilon_2(\delta)]}
\\
&=2^{-n[H(X_1)-H_d(X_1|Y)+H(X_2|\overline{V})-H_d(X_2|Y,\overline{V})-\epsilon_1(\delta)-\epsilon_2(\delta)-\epsilon_3(\delta)-\epsilon_4(\delta)]}.
\end{flalign}

Therefore  $\Pr(\mathcal{E}_5)\rightarrow 0$ as $n\rightarrow\infty$ if
\begin{flalign}
R_1+R_2 <I_d(X_1;Y)+I_d(X_2;Y|\overline{V}).
\end{flalign}

Further, by the packing lemma \cite[p. 46]{NetworkInformationTheory}  $\Pr(\mathcal{E}_6)$ tends to $0$ and $n$ tends to infinity if
\[R_2 < I_d(X_2;Y|\overline{V}).\]

Since the encoder does not know the delay  $d\in\mathcal{D}$,  a rate-pair is achievable for fixed $P(x_1),P(x_2|\overline{v})$ if it lies in the intersection of all the regions $\underline{R}_d(P(x_1),P(x_2|\overline{v}))$. Therefore,
\begin{flalign}
\underline{\mathcal{R}}(P(x_1),P(x_2|\overline{v}))&=\bigcap_{d\in\mathcal{D}}
\underline{\mathcal{R}}_d(P(x_1),P(x_2|\overline{v}))\nonumber\\
& =\left\{
(R_1,R_2):\begin{array}{ll}
    \multirow{2}{*}{}&\hspace{1cm}R_1\leq H(X_1),\\
    &\hspace{1cm}R_2 \leq \min_{d\in\mathcal{D}}I_d(X_2;Y|\overline{V}),\\
    & R_1 +R_2 \leq \min_{d\in\mathcal{D}}\left[I_d(X_1;Y)+I_d(X_2;Y|\overline{V})\right]
\end{array}
\right\}.
\end{flalign}
where the last equality follows sice the set of all possible delays is finite.

Consequently, the following rate region
\begin{flalign}
\underline{\mathcal{R}}&=\bigcup_{P(x_1),P(x_2|\overline{v})}\underline{\mathcal{R}}(P(x_1),P(x_2|\overline{v})),
\end{flalign}
is achievable.

Finally, since $d_{max},d_{min}<\infty$ we can use time sharing arguments to infer that the closure convex of the rate region $\underline{\mathcal{R}}$ is an achievable rate region for the ACMAC.

\bibliographystyle{IEEEtran}

\end{document}